\documentclass[acmtog]{acmart}

\renewcommand\footnotetextcopyrightpermission[1]{}

\usepackage{array}
\usepackage{gensymb}
\usepackage{booktabs}
\usepackage{multirow}
\usepackage{makecell}
\usepackage{color}
\usepackage{bm}

\def\bwi{\bm{\omega}_i}
\def\bwo{\bm{\omega}_o}
\def\bv{\bm{V}}

\def\myfigure#1#2#3{\begin{figure}[htb]\centering\includegraphics[width = \linewidth]{#2}\caption{#3}\label{fig:#1}\end{figure}}
\def\mycfigure#1#2#3{\begin{figure*}[htb]\centering\includegraphics*[clip, width = \linewidth]{#2}\caption{#3}\label{fig:#1}\end{figure*}}

\usepackage{pifont}
\newcommand{\cmark}{\ding{51}}
\newcommand{\xmark}{\ding{55}}
\usepackage[ruled]{algorithm2e}

\SetAlFnt{\small}
\SetAlCapFnt{\small}
\SetAlCapNameFnt{\small}
\SetAlCapHSkip{0pt}

\acmJournal{TOG}

\begin{document}
\title{Neural BRDFs: Representation and Operations}

\newcommand{\added}[1]{\textcolor{red}{#1}}

\author{Jiahui Fan}
\affiliation{
    \institution{School of Computer Science and Engineering, Nanjing University of Science and Technology}
    \country{China}
}
\email{fjh@njust.edu.cn}

\author{Beibei Wang$^\dagger$}
\thanks{$^\dagger$Corresponding authors. Email: beibei.wang@njust.edu.cn}
\affiliation{
    \institution{School of Computer Science and Engineering, Nanjing University of Science and Technology}
    \country{China}
}
\email{beibei.wang@njust.edu.cn}

\author{Milo\v{s} Ha\v{s}an}
\affiliation{
    \institution{Adobe Research}
    \country{USA}
}

\author{Jian Yang$^\dagger$}
\thanks{$^\dagger$Corresponding authors. Email: csjyang@njust.edu.cn}
\affiliation{
    \institution{School of Computer Science and Engineering, Nanjing University of Science and Technology}
    \country{China}
}
\email{csjyang@njust.edu.cn}

\author{Ling-Qi Yan}
\affiliation{
    \institution{University of California, Santa Barbara}
    \country{USA}
}
\email{lingqi@cs.ucsb.edu}

\begin{abstract}
Bidirectional reflectance distribution functions (BRDFs) are pervasively used in computer graphics to produce realistic physically-based  appearance. In recent years, several works explored using neural networks to represent BRDFs, taking advantage of neural networks' high compression rate and their ability to fit highly complex functions. However, once represented, the BRDFs will be fixed and therefore lack flexibility to take part in follow-up operations. In this paper, we present a form of ``Neural BRDF algebra'', and focus on both representation and operations of BRDFs at the same time. We propose a representation neural network to compress BRDFs into latent vectors, which is able to represent BRDFs accurately. We further propose several operations that can be applied solely in the latent space, such as layering and interpolation. Spatial variation is straightforward to achieve by using textures of latent vectors. Furthermore, our representation can be efficiently evaluated and sampled, providing a competitive solution to more expensive Monte Carlo layering approaches.
\end{abstract}



\begin{teaserfigure}
    \centering
    \includegraphics[width=\textwidth]{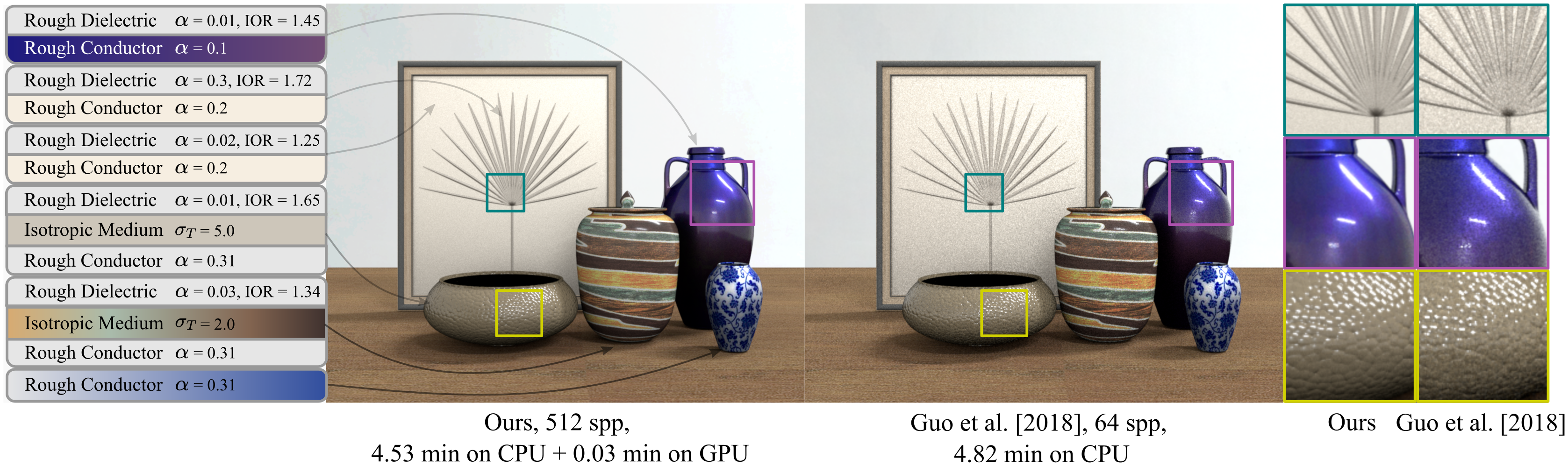}
    \caption{We present a neural representation for BRDFs; the latent vectors of this representation support several operations including layering. Our method is able to produce closely matching layered results to the Monte Carlo simulation of Guo et al.~\shortcite{Guo:2018:Layered} with less cost, and works well with spatially-varying parameters (here we show varying top layer roughness, IOR, bottom layer Fresnel value, macro surface normal and scattering medium properties).}
    \label{teaser}
\end{teaserfigure}

\maketitle

\section{Introduction}
\label{sec:intro}

Spatially-varying bidirectional reflectance distribution functions (SVBRDFs) are pervasively used in computer graphics to produce vivid and realistic appearance. However, no single BRDF model has succeeded in satisfying all requirements in different scenarios. Analytic models, such as microfacet BRDFs with Beckmann and GGX normal distributions, have computational efficiency but can produce less realistic appearances, due to the discrepancy between the microfacet assumptions and the real-world light-matter interactions. On the other hand, measured SVBRDFs and bidirectional texture functions (BTFs) can faithfully recover the real-world appearance, but have expensive storage and lack flexibility; that is, once measured, these BRDFs are generally fixed. Layering simple analytic models into more complex composites is an attractive alternative explored in recent years, though the rendering solutions for layered materials can be mathematically complex and/or computationally expensive.

In recent years, neural networks for representing SVBRDFs and BTFs has received attention \cite{Rainer2019Neural, Rainer2020Unified, kuznetsov2021neumip, takikawa2021neural}. These neural approaches mostly focus on efficient compression \cite{Rainer2019Neural, Rainer2020Unified, takikawa2021neural} and query \cite{kuznetsov2021neumip}. They have greatly reduced the storage overhead, successfully making high-dimensional SVBRDF and BTF data practically usable in rendering. However, these methods are primarily different ways of representing and compressing measured BRDFs; in our opinion, more operations need to be supported on the compressed representation to make it truly useful in practice.

Which operations are the most desirable? In a modern rendering framework with multiple importance sampling (MIS), any BRDF must support not only \emph{evaluation}, answering a point query for a given position and view/light directions, but also \emph{importance sampling}, distributing outgoing direction samples according to the 2D slice of a BRDF given the incoming direction. Further, BRDFs are often mixed/interpolated to generate new appearance, texture mapped to specify properties per spatial location, and layered to introduce coatings and other effects.

Our approach is to first compress BRDFs into short latent vectors. Inspired by recent neural approaches that operate on the latent space for geometry transformations \cite{granskog2021neural}, we train additional networks that operate solely in the latent space, providing individual operations to BRDFs, such as layering. From the outside, these neural networks hide the actual implementation of such operations as if they are the generic operators and our compressed BRDF are operands, leading to a form of ``Neural BRDF algebra''.

To achieve spatial variation, we simply use a multi-channel texture (with each texel holding a latent vector) to define SVBRDFs. This SVBRDF map can be efficiently rendered, or combined with other BRDF maps for interpolation and layering to create new SVBRDFs. We show several examples of this black-box style use of our neural operators, as well as demonstrate faster performance compared to previous Monte Carlo techniques.

\section{Related Work}
\label{sec:related}

\paragraph{Traditional SVBRDF/BTF compression.}  Due to the high dimensionality of BTFs~\cite{Dana:1999:BTF}, traditional approaches compress BTF, using principal component analysis (PCA)~\cite{Weinmann:2014:PCA}, linear matrix factorization techniques~\cite{Koudelka03compression,Sattler:2003:Cloth,
KIM2018:Compress} or hierarchical tensor decomposition methods~\cite{Wang:2005:Tensor, ruiters-2009-ksvd}. 
None of these compressing approaches are able to compress a large group of BTFs and lack the ability to operate on the compressed data.

\paragraph{Neural SVBRDF/BTF compression.}
Kuznetsov et al.~\shortcite{kuznetsov2021neumip} introduce a neural method for representing and rendering a variety of material appearances at different scales. In their method, one neural network is trained per material. More recently, Sztrajman et al.~\shortcite{Sztrajman:2021:Neural} represent each BRDF with a decoder structure, resulting in better quality, at the cost of more storage for each BRDF. We categorize the aforementioned approaches as \emph{specialized methods}, because one neural network only represents one material/BRDF in these works. The specialized methods are usually of high quality, but their representation is usually costly to store, and is especially difficult for operations, since the operations will have to take neural networks as inputs and outputs.

The other kind of approaches are \emph{generalized methods}. Rainer et al.~\shortcite{Rainer2019Neural} proposed a neural representation for BTFs, where each per-texel BRDF is represented with a latent vector, resulting in a compact representation of the BTFs. Strictly speaking, this approach is still not fully generalized, since it requires training an autoencoder architecture per BTF and does not generalize across materials. Later, Rainer et al.~\shortcite{Rainer2020Unified} extended the work and introduced a unified model to represent all the materials. The generalized methods are more friendly to operations, thanks to the fixed pipeline that converts general BRDFs into the latent space. But as a trade-off, they are often of lower quality. Specifically, neither Rainer et al.~\shortcite{Rainer2019Neural} nor Rainer et al.~\shortcite{Rainer2020Unified} work well for highly specular materials. Compared to these two works, our method also represents the BRDFs with latent vectors, but our method provides better quality, especially for highly specular BRDFs, thanks to the design of our representation network.

Being a generalized method, our method achieves comparable representation quality as compared to specialized methods, such as Sztrajman et al.~\shortcite{Sztrajman:2021:Neural}, but it requires much less storage, making it more suitable for SVBRDFs' representation. More importantly, our method not only represents BRDFs with latent vectors, but also supports a rich set of operations on the latent vectors.

\paragraph{Other related neural approaches.} 
Beside BRDF representation, neural networks have also been used for other rendering applications (e.g., glints computation, multiple scattering representation in participating media, etc.) Kuznetsov et al.~\shortcite{Kuznetsov19:GlintsGAN} propose a generative adversarial model for generalized normal distribution function (GNDF) representation, which avoids the run-time glints computation and texture synthesis. However, the model has to be trained for different groups of normal maps / height fields. Ge et al.~\shortcite{Ge:2021:multiple} represent the multiple scattering for the entire homogeneous participating media space with a neural network. Zhu et al.~\shortcite{Zhu:2021:Luminaires} compress a complex luminarie's light field into an implicit neural representation, which enables efficient BRDF evaluation, importance sampling and probability density function (pdf) computation.

Recently, Mildenhall et al.~\shortcite{Mildenhall2020NeRFRS} learn a radiance field representation (NeRF) via differentiable volume rendering, which inspired many follow-up works. These methods are focused on object capture and do not generally consider materials as separate components, with a few exceptions, e.g., Bi et al.~\shortcite{Bi:2020:Appearance}.
Recent neural approaches also propose to operate on the latent space for geometry transformations~\cite{granskog2021neural}; our method takes a similar approach to materials.

More applications of neural networks in rendering focus on specific sub-problems in other parts of the material modeling and/or light transport processes. They include function mapping~\cite{yan2017furbssrdf} which maps fur parameters to participating media, path guiding~\cite{Muller:2019:sampling} by learning a radiance distribution and learning screen space special effects for buffers~\cite{Nalbach:2017:DeepShading} and neural texture~\cite{Thies:2019:Neural} for deferred rendering.

\paragraph{BRDF layering operations.} Layering is an important operation for BRDFs, which has been addressed by several lines of work, including approximate analytic models \cite{Weidlich:2007:layering}, Fourier basis functions \cite{jakob:2014:layered,Jakob2015Layerlab,Zeltner2018Layer}, Monte Carlo simulation based approaches (\cite{Guo:2018:Layered}, \cite{Gamboa:2020:EfficientLayered}, and \cite{Xia:2020:Layered}) and tracking directional statistics (\cite{Belcour:2018:Layered},\cite{Yamaguchi:2019:anisotropic}, and \cite{ WeierAndBelcour:2020:Anisotropic}). The Fourier-based methods rely on expensive computation per parameter setting, requiring many coefficients especially for low-roughness surfaces, which makes it difficult to handle spatially-varying textures. The Monte Carlo based methods are able to produce high-quality results, and support spatially-varying textures, thus we treat them as ground-truth. However, the required random walks lead to extra variance (noise) added to the rendered results. The third group of works expresses the directional statistics (e.g., mean and variance) of a layered BRDF and track the statistical summary at each step, resulting in high performance, even achieving real-time frame rates. However, summary statistics are a fundamentally approximate way of representing the underlying functions.

Compared to all of these works, our method represents the BRDF with a latent vector, and performs the operations on the latent vectors. While supporting other rendering-related operations, such as importance sampling, we specifically treat the layering operation as a complex and challenging task, demonstrating the ability of our black-box operand-operator style approach. Since we use Monte Carlo based approaches as the ground-truth for training, our layering results are very close to Monte Carlo based approaches, while our method avoids the expensive random walk, and does not introduce additional variance (noise).

\section{Neural BRDFs and Operations}
\label{sec:method}

In this section, we present our solution to neural BRDF representation and corresponding operations. We first formulate the problem (Sec.~\ref{subsec:formulation}). Then, we analyze the requirements of our Neural BRDF representation, leading to our general-purpose BRDF decoder structure (Sec.~\ref{subsec:representation}). Finally, we introduce individual neural networks for different neural operations.

\subsection{Overview and formulation}
\label{subsec:formulation}

We focus on representing individual BRDFs and providing neural operations on them. This is key to our design offering flexibility, and it differentiates our method against previous work that compresses the entire chuck of SVBRDFs/BTFs~\cite{Rainer2019Neural}.

A BRDF is a 4D function $f(\bwi, \bwo)$, where $\bwi$ and $\bwo$ are the incoming and outgoing directions on the unit hemisphere. Note that this definition can be extended to bidirectional scattering distribution functions (BSDFs) by considering full unit spheres for directions. For simplicity, we do not represent BSDFs in this paper. However, we do implicitly consider BSDFs when layering one BRDF atop another: the top BRDF is assumed to transmit all energy that is not reflected, and this affects the layering operation.

We will start from compressing any BRDF in a compact neural form:

\vspace{0.1in}\noindent\fbox{\begin{minipage}{0.975\columnwidth}\vspace{0.05in}
\textbf{Representation:}\vspace{0.05in}
\begin{equation}
    f(\bwi,\bwo) \xrightarrow{N_{\textrm{rep}}}
 \bv_f, 
\end{equation}\vspace{0.02in}
\end{minipage}}\vspace{0.1in}
where $\bv_f$ is known as a latent vector and $N_{\textrm{rep}}$ is a neural representation projecting operator. As we will see later, this operator is implemented through optimization (searching for a latent vector that decodes to the input BRDF).

The representation should be \emph{general-purpose}, taking in any BRDF as input, outputting its corresponding latent vector. In other words, it is not sufficient to train one different network for each BRDF \cite{Sztrajman:2021:Neural}, or even one network per SVBRDF \cite{Rainer2019Neural}.

Once the BRDFs are represented as latent vectors, we treat them as operands, and provide operators that act upon them. Specifically, we focus on these operations:

\vspace{0.1in}\noindent\fbox{\begin{minipage}{0.975\columnwidth}\vspace{0.05in}
\textbf{Evaluation:}\vspace{0.05in}
\begin{equation}
    \{\bv_f,\bwi,\bwo\} \xrightarrow{N_{\textrm{eval}}} f(\bwi,\bwo),
\end{equation}\vspace{0.05in}

\textbf{Interpolation (blending):}\vspace{0.05in}
\begin{equation}
    \{\bv_{f_i},w_{i}\}
      \xrightarrow{N_{\textrm{interp}}} \sum_i w_{i}\bv_{f_i},
\end{equation}\vspace{0.05in}

\textbf{Importance sampling:}\vspace{0.05in}
\begin{equation}
    \{\bv_f,\bwi\} \xrightarrow{N_{\textrm{sample}}} \bwo\sim f(\bwi,\cdot),
\end{equation}\vspace{0.05in}

\textbf{Layering:}\vspace{0.05in}
\begin{equation}
    \{\bv_{\textrm{top}},\bv_{\textrm{bottom}}, A, \sigma_T \} \xrightarrow{N_{\textrm{layering}}} \bv_{\textrm{layered}}.
\end{equation}
\vspace{0.02in}
\end{minipage}}\vspace{0.05in} \\

Above, $w_i$ represents the weights for the $i^\textrm{th}$ latent vector, and $\bv_\textrm{top}$ and  $\bv_\textrm{bottom}$ represent the latent vectors for BRDFs at the top layer and the bottom layer respectively. $A$ and  $ \sigma_T $ are the single-scattering albedo and extinction coefficients for the participating medium inserted between the two layers.

These operations cater to the way BRDFs are used in rendering applications. For example, BRDF evaluation comes in the form of point queries, requesting one BRDF value given a pair of incoming and outgoing directions $\bwi$ and $\bwo$ on a given shading point. Importance sampling requires us to find an efficient (but not necessarily exact) way of sampling an outgoing direction according to the shape of the 2D outgoing BRDF slice given the incoming direction.

Also note that different operations may have different computational cost. The interpolation operation $N_\textrm{interp}$ is a simple linear blend in latent space and does not require a neural network, while the layering operation $N_\textrm{layering}$ can be complex.

\subsection{Neural BRDF representation and evaluation}
\label{subsec:representation}

\myfigure{network_general}{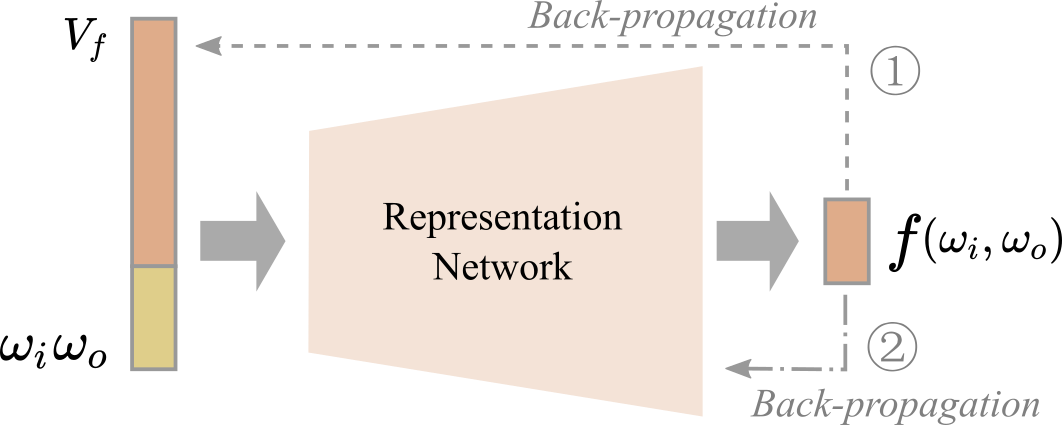}{The high-level architecture of the evaluation network. The network only includes a decoder, which is used for both BRDF evaluation and representation. When we train the network, we back-propagate the gradients through routes \textcircled{1} and \textcircled{2}. When we need the representation of any new material, we freeze the network's parameters and use only the back propagate route \textcircled{1} to optimize the latent vectors.} 

\mycfigure{representation_network_detail}{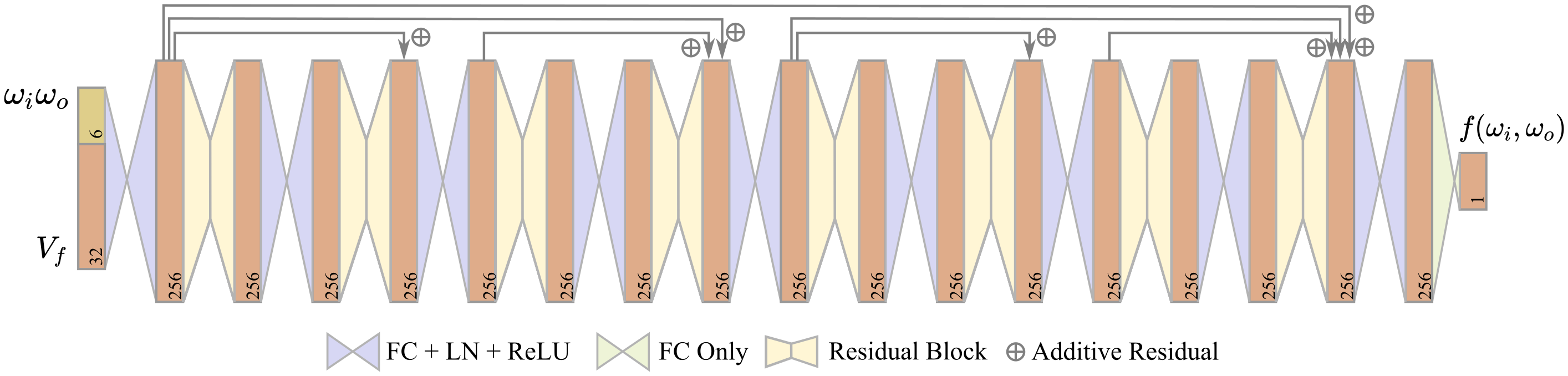}{The structure details of the evaluation network. The purple connection denotes an FC+LN+ReLU combination, while the green one is FC only. The yellow module refers to the residual block. The dark gray arrows mean addition skip connections between layers. All residuals and skip connections are added before the normalization and activation layers. The dimensions of the feature vectors are marked as black numbers in the figure. The detailed structure of a residual block will be illustrated in Figure~\ref{fig:layering_network}.}

We would like to find a general-purpose neural network that is able to compress any input BRDF $f(\bwi,\bwo)$ into a latent vector $\bv_f$. We do not want to pre-specify the discretization of the input BRDF. Instead, we opt to define an \emph{evaluation} network architecture, which takes a latent vector as well as incoming and outgoing directions as input, and returns the corresponding BRDF value. To project a BRDF into the latent space, we simply optimize for a latent vector that gives back the input BRDF at any desired discretization.

We design the architecture of our Neural BRDF evaluation network as shown in Figure~\ref{fig:network_general}. It takes a latent vector and an incoming-outgoing pair as query, and outputs the corresponding BRDF value. The network is trained by back-propagation on a dataset of BRDFs, as detailed later.

Note that we have two back-propagation routes, one for the latent vector, the other for the weights of the evaluation network. When we are training the network, we use all values of BRDFs across the dataset, and we back-propagate the gradient through both routes, updating the weights in the network and the latent vectors simultaneously. In this way, our network learns to use different latent vectors to represent different BRDFs. Meanwhile, all latent vectors are interpreted the same way through the evaluation network. On the other hand, once we have trained the evaluation network and would like to use it to project any new BRDF into the latent space, we freeze the network parameters and only back-propagate to the latent vector. For BRDFs with different roughness, this projection takes from less than $10$ seconds to $45$ seconds to converge on an RTX 2080Ti GPU.

Figure~\ref{fig:representation_network_detail} illustrates the detailed architecture of our evaluation network. Our network is a multi-layer perceptron (MLP), with each fully connected (FC) layer followed by a layer normalization (LN) and ReLU activation (except the output layer). We use a bottleneck residual structure as our basic module in our network, which has one hidden layer with a half amount of units and residual from the first layer to the last. We alternatively use the residual blocks and simple FC+LN+ReLU layers to build our network, and add several skip connections. This kind of combination is repeated for eight times in total. The basic number of hidden units is set to $256$. The dimension of the latent vector is set to be $32$. We treat every channel of the RGB color space separately, and we do not clamp any large high dynamic range (HDR) values.

\begin{table}
    \centering\setlength{\tabcolsep}{2pt}
    \caption{Comparison of different BRDF representation methods, considering the ability for representing specularity, SVBRDFs and the generality of the model. The generality means the ability to represent all the materials with a single network.}
    \begin{tabular}{cccc} 
        \toprule
        Method & Specularity & SVBRDFs & Generality\\
      \midrule
      Rainer et al.~\shortcite{Rainer2019Neural}&       \xmark  &  \cmark   &  \xmark \\
      Rainer et al.~\shortcite{Rainer2020Unified}&      \xmark  &  \cmark   &  \cmark \\
     Sztrajman et al.~\shortcite{Sztrajman:2021:Neural} & \cmark  & \xmark  &  \xmark \\
      \textbf{Ours}                             &          \cmark  &  \cmark  & \cmark \\
      \bottomrule
      \end{tabular}
      \label{tab:comparison}
\end{table}

With our evaluation network, any BRDF can be compressed into a latent vector. We do not require the BRDFs to be parametric or measured, nor do we care whether they are already layered or not. We do however make two assumptions for simplicity. First, we assume that all BRDFs are isotropic for now, mainly for the efficiency of the training dataset. Second, we assume that none of the BRDFs are normal mapped, because this operation is easier to achieve by altering local shading coordinates during rendering.

Our evaluation network can successfully represent BRDFs from commonly seen materials. We plot and render with some latent vectors in Figure~\ref{fig:rep_matpreview} using our evaluation operator introducing soon. In addition to representing a single BRDF, we are able to define a \emph{latent texture}, where each texel is a latent vector representing a BRDF. In this way, we can use this latent texture to describe SVBRDFs, as will be demonstrated on more examples in Sec.~\ref{sec:result}.

\textbf{Discussion.} In Table~\ref{tab:comparison}, we compare our method against other three related works, regarding the representation accuracy for specularity, ability for representing SVBRDFs, and the generality of the model. Regarding representation accuracy, both Sztrajman et al.~\shortcite{Sztrajman:2021:Neural} and our method are able to handle sharp specular BRDFs, while Rainer et al.~\shortcite{Rainer2019Neural} and Rainer et al.~\shortcite{Rainer2020Unified} are less accurate. However, Sztrajman et al.~\shortcite{Sztrajman:2021:Neural} require a decoder for each BRDF representation, which makes it not applicable for SVBRDF representation. Regarding the generality, Sztrajman et al.~\shortcite{Sztrajman:2021:Neural} require training for each BRDF, and Rainer et al.~\shortcite{Rainer2019Neural} require training for each single SVBRDF.

\subsection{Neural BRDF interpolation and mipmapping}
\label{subsec:interp}

\myfigure{interpolation}{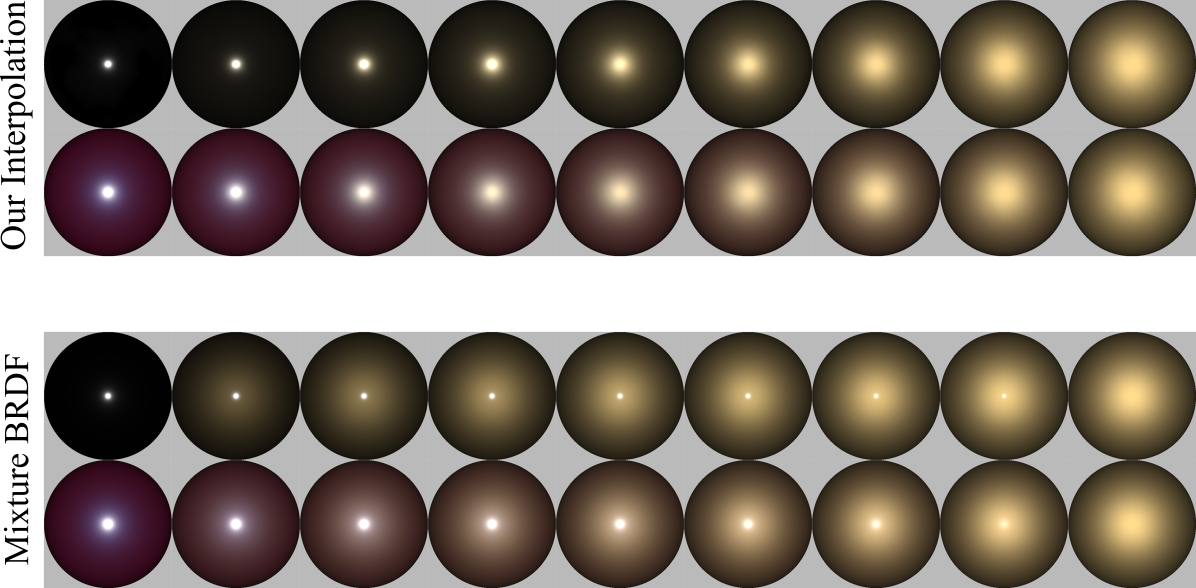}{Comparison between our latent space interpolation and linear interpolation (blending/mixture of BRDFs). Our method produces a more natural transition from low roughness to medium roughness then high roughness. But the mixture of BRDFs always keeps two lobes and the highlight remains sharp, which is unnatural.}

\myfigure{mipmap}{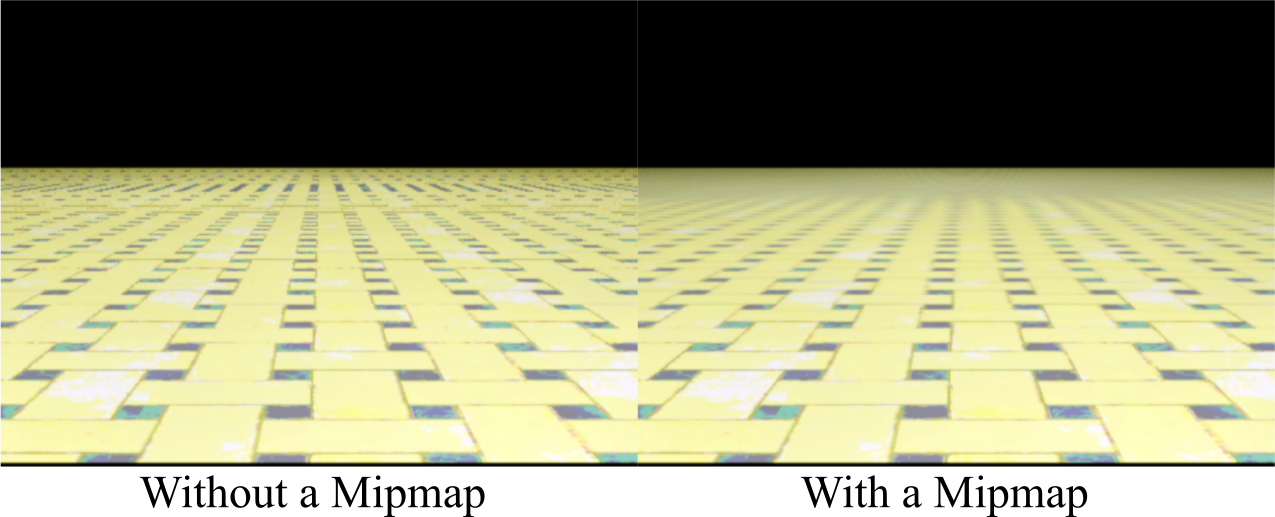}{Using our method without and with latent texture mipmapping, both rendered at 1 spp. The mipmap is generated from the multi-channel latent BRDF texture as a precomputation, and is queried on the fly in the standard way, using the appropriate level with trilinear interpolation. The mipmap reduces aliasing even with very low sampling rates.}

By interpolating two or more given BRDFs, a new BRDF can be obtained, showing a natural transition effect between the input materials. We interpolate the BRDFs by performing the interpolation of the latent vectors of two input BRDFs. Figure \ref{fig:interpolation} shows a visualization of our interpolation, compared to naive linear interpolation of the BRDF values themselves.

As mentioned earlier, SVBRDFs are represented as latent textures, where each texel specifies a latent vector. When querying the latent texture, we perform bilinear interpolation on the latent vectors. Furthermore, we support building a mipmap for the latent texture, where texels from higher levels are computed by averaging the latent vectors of four texels from the previous level. During rendering, we compute the pixel's footprint for each shading point, and then query the mipmapped latent texture with trilinear interpolation, using the footprint size to find a proper level in the mipmap. Figure \ref{fig:mipmap} shows the results of this interpolation are as expected.

\subsection{Neural BRDF importance sampling}
\label{subsec:sample}

\myfigure{sampling_compare}{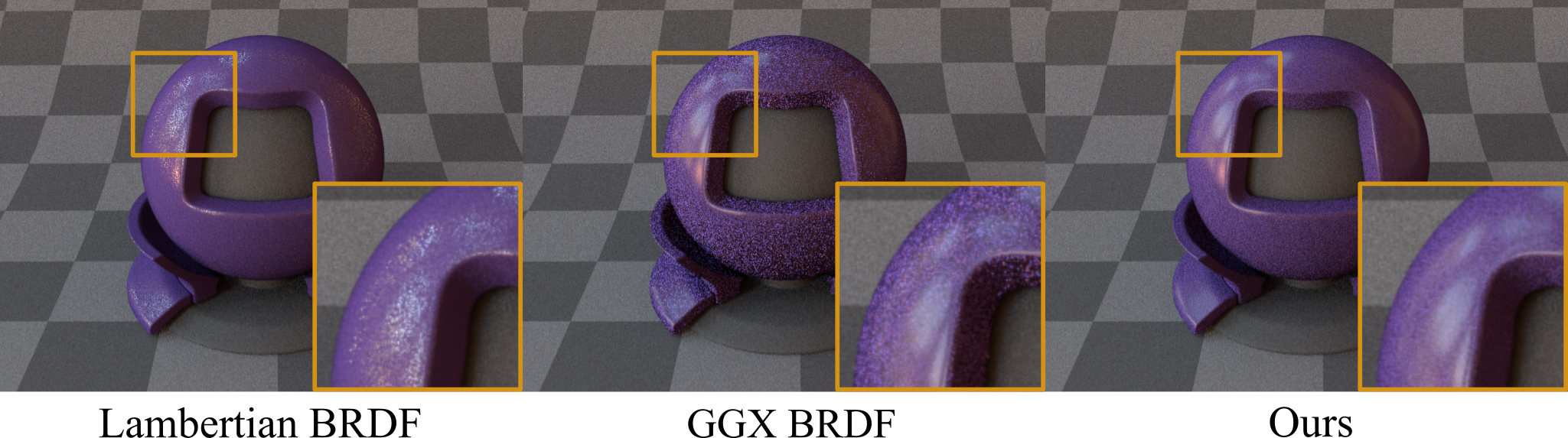}{Comparison between three sampling strategies. Left: sampling the outgoing Lambertian lobe. Middle: sampling according to the GGX lobe with parameters obtained from the top layer. Right: our method, sampling two lobes predicted by our sampling network. All results are produced using BRDF sampling only, with 256 spp. Our method has the least variance.}

\myfigure{MIS_validate}{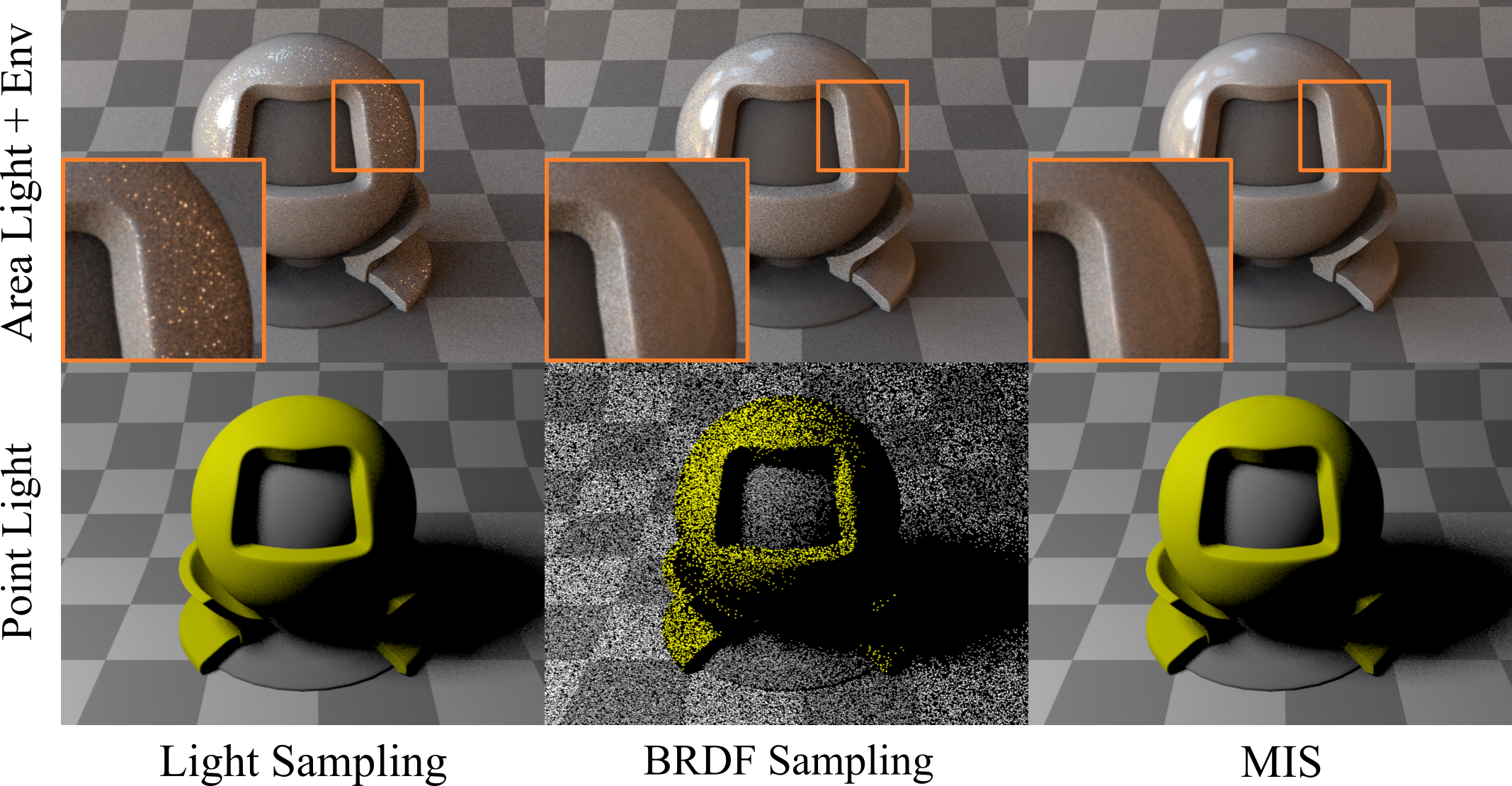}{Our method can be naturally applied in the MIS framework. We compare between light sampling only, BRDF sampling only and MIS under two different lighting configurations (large light sources vs. a small light source). In both cases, MIS produces the best results, as expected.}

\mycfigure{layering_network}{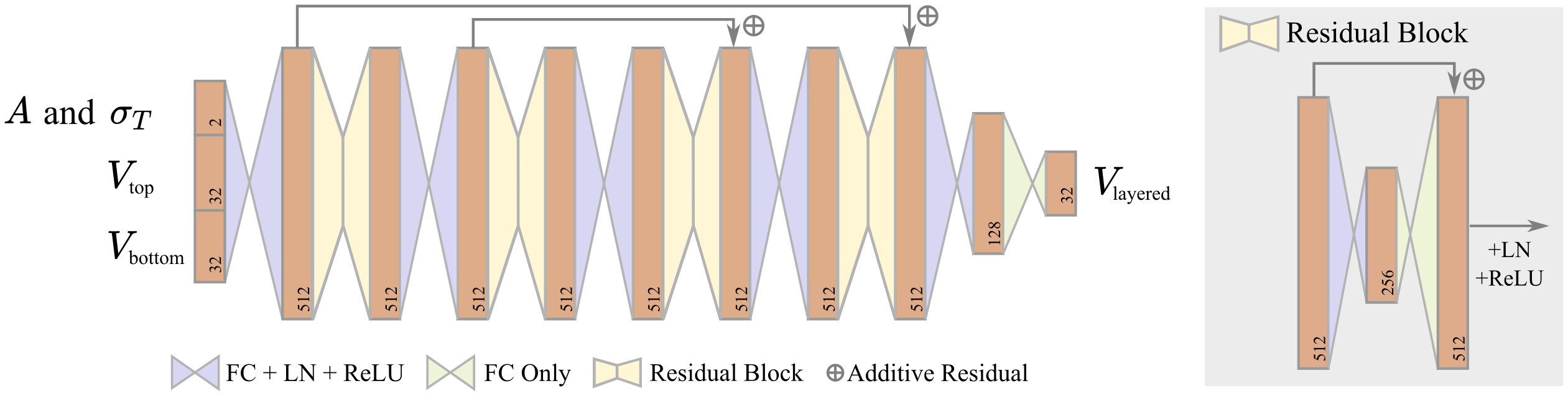}{The structure of layering network. $A$ and $\sigma_T$ are scalars that represent the albedo and the extinction coefficients. $\bv_{\textrm{top}}$ and  $\bv_{\textrm{bottom}}$ are the latent vectors to represent the top layer and the bottom layer respectively. The rest of notations are the same as Figure~\ref{fig:representation_network_detail}.}

Importance sampling is a critical operation for including a BRDF in a practical path tracing system. Specifically, for a given incoming direction, we want to choose an outgoing direction with a pdf roughly proportional to the outgoing BRDF lobe as a function on the hemisphere; we also need to be able to evaluate the sampling pdf for a given direction. To introduce a sampling operation for Neural BRDFs represented in our latent space, our approach is to use an analytic proxy distribution to mimic the actual BRDF lobe. We currently use a weighted sum of a Gaussian lobe and a Lambertian lobe; this approach could be easily extended to include multiple Gaussian lobes if needed.

Our pdf proxy is defined as
\begin{equation}
    \textrm{pdf}(\bwi,\bwo) = (1 - w) G_\sigma(h_x, h_y) + w L(\bwo)
    \label{eq:pdf}
\end{equation}
where $(h_x, h_y)$ represents the projected half vector (its $z$-component is dropped), and $G_\sigma$ is a Gaussian function with standard deviation $\sigma$, normalized on the projected hemisphere. $L$ is the Lambertian pdf on the outgoing hemisphere (i.e., $\cos \theta_o / \pi$), where $\theta_o$ is the angle between the outgoing direction $\omega_o$ and the macro surface normal.

To obtain the parameters, $\sigma$ for the Gaussian function and $w$ for the weight, we propose a sampling network to learn these parameters. To do this, we define the concept of a \emph{generalized normal distribution function} (GNDF) of the BRDF. The name is chosen because for a microfacet BRDF, its GNDF has a similar (though not identical) shape to its NDF. The GNDF is the normalized average of the BRDF, in half-angle space, over all incoming directions $\omega_i$; it is thus a 2D function of the half-vector. In practice, we estimate the GNDF by uniformly sampling $40 \times 40$ incoming vectors on the upper hemisphere and averaging the resulting 2D BRDF lobes over half-angle space. 

We train the importance sampling network by minimizing the difference between our pdf proxy (Equation~\ref{eq:pdf}) and the GNDF, in the following sense. Our sampling network is a simple four-layer MLP (with 128, 512, 128 and 32 hidden units individually). It takes any BRDF latent vector and an incoming direction as input, and outputs the sampling parameters $\sigma$ and $w$ of this BRDF. For each latent vector, we generate $40 \times 40$ different incoming directions, then we take the averaged parameters and the averaged $\textrm{pdf}(\bwi,\bwo)$ predicted by the network from these individual incoming directions. Finally, we match the $\textrm{pdf}$ and the GNDF using a loss to update our network weights.

When we sample a BRDF according to our proxy Equation~\ref{eq:pdf}, we firstly generate a random number to choose between the Lambertian and the Gaussian components, with the probability of diffuse ratio $w$. Then we importance-sample the chosen component to obtain the outgoing direction $\omega_o$ (in case of the Gaussian lobe, this is done by first sampling the half vector $h$ and transforming it into outgoing direction). We finally calculate the pdf value for the chosen outgoing direction; this uses the Jacobian term of the half-angle transform, as detailed by Walter et al. \shortcite{Walter07}. We validate the results of our sampling methods in Figures~\ref{fig:sampling_compare} and \ref{fig:MIS_validate}.

Note that by definition, our pdf proxy only depends on $\sigma$ and $w$. Hence, even though the importance sampling takes an incoming direction as input, the pdf proxy parameters are independent of it, and can be computed only once using our sampling network for a given BRDF before rendering. Therefore, our importance sampling is very fast, because no network inference is be performed on the fly.

\subsection{Neural BRDF layering}
\label{subsec:layering}

\myfigure{layer_config}{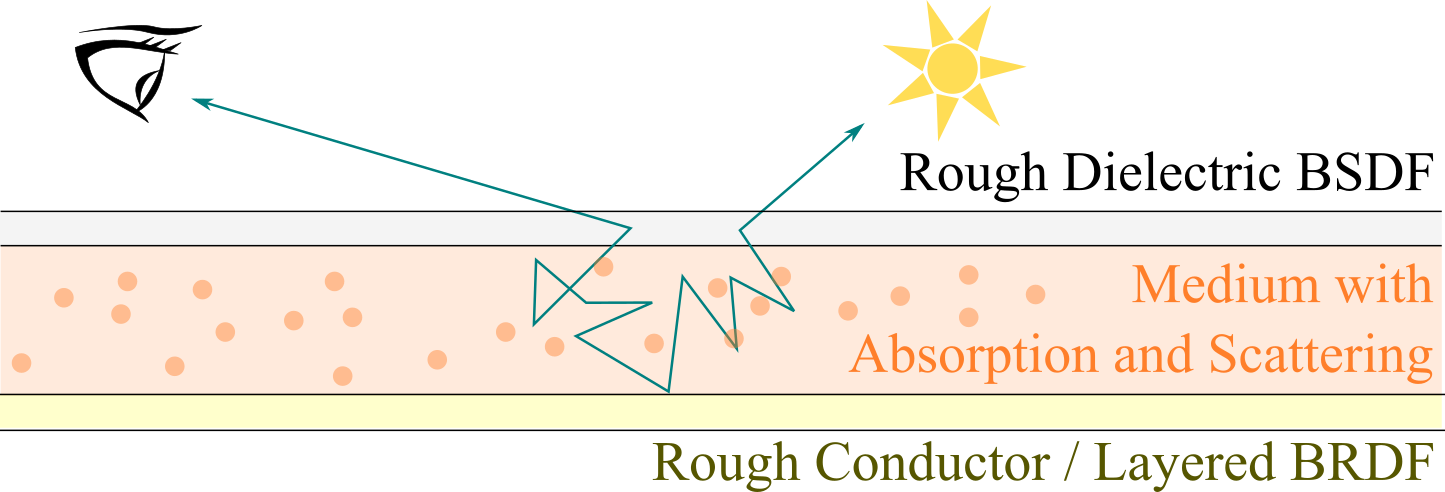}{Our configuration for layered BRDFs includes a top layer using a rough dielectric, a bottom layer using a rough conductor or another layered BRDF and a homogeneous participating media in the middle. }

Layering BRDFs (Figure~\ref{fig:layer_config}) includes complex light transport interactions among the layers. The most accurate way to layer BRDFs is using a Monte Carlo random walk \cite{Guo:2018:Layered}; however, this is expensive, especially when there are dense volumetric media between the interfaces; the random walk also introduces variance.

We instead propose to learn the layering operation in the latent space. We only consider a two-layer operation, consisting of a top layer with a rough dielectric BSDF (whose transmission component is implied as the energy complement of the top BRDF, with no energy lost in the interface itself) and a bottom layer with any BRDF, and a layer of homogeneous participating media with albedo $A$ and extinction coefficient $\sigma_T$ (assuming isotropic scattering) between the interfaces, and we do not use the extra multiple scattering component that compensates for energy loss in advanced micro-facet models. Note that this two-layer setup can be easily extended into a multi-layered configuration by recursively applying it, as we discuss in Sec.~\ref{sec:quality}.

Thanks to the generality of our BRDF latent space, the layered BRDF can also be represented with the latent vector. Thus, the goal of layering is getting a latent vector which represents the layered BRDF, from the input configurations. We propose a layering network to find the mapping between the input component BRDFs and the layered BRDF, as shown in Figure~\ref{fig:layering_network}. The network has a similar structure with the evaluation network, with fewer layers (stack for only four times) but more hidden units each layer (512 basically). The layering network takes two latent vectors which represents the top layer and bottom layer as input, together with the albedo and extinction coefficients. It directly outputs one target latent vector to represent the layered BRDF. Similar to the evaluation network, our layering network also deals with RGB channels independently.

\section{Implementation Details}
\label{sec:implementation}

\mycfigure{rep_matpreview}{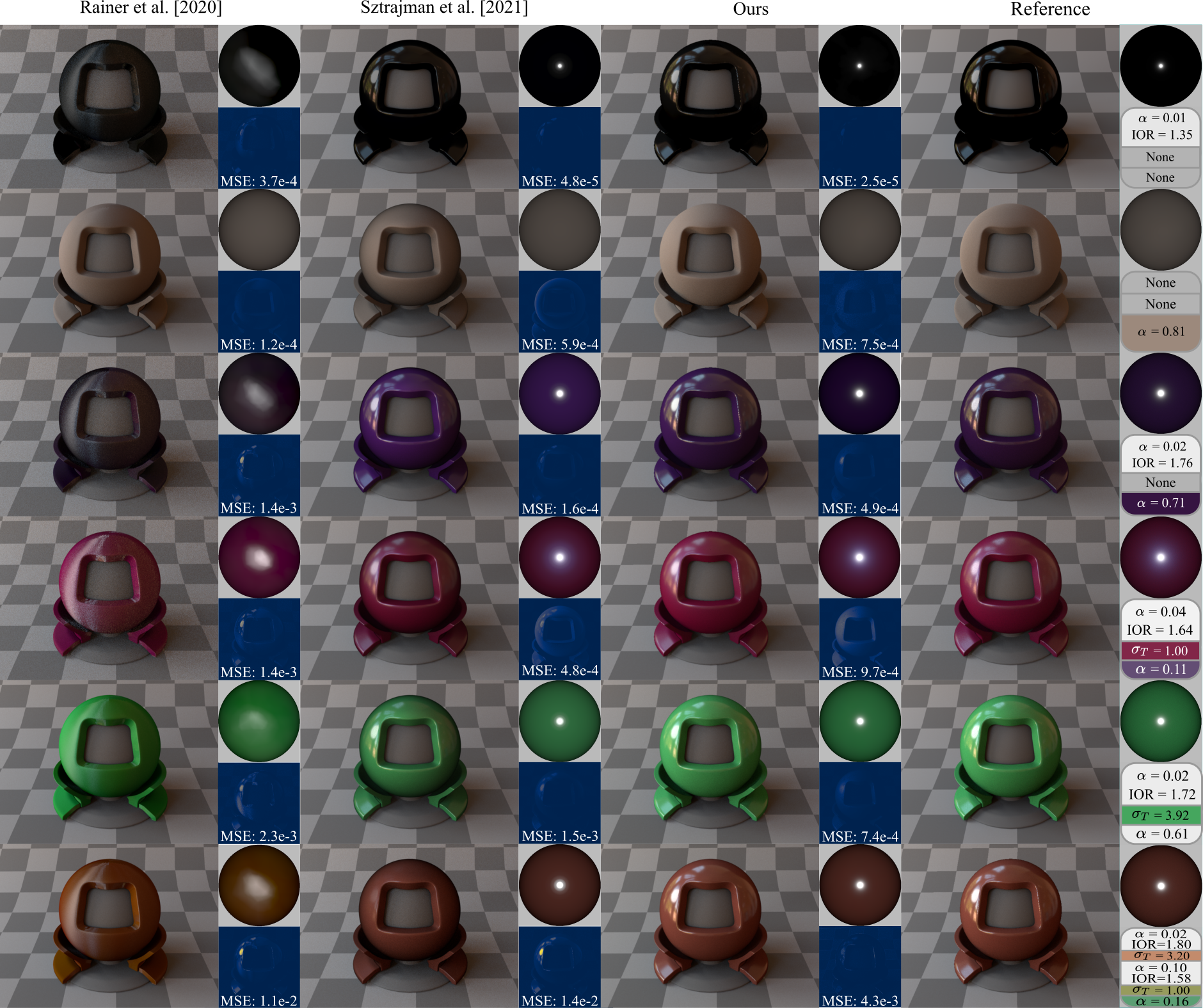}{Comparison of the representation abilities with different BRDF configurations between Rainer et al.~\shortcite{Rainer2020Unified}, Sztrajman et al.~\shortcite{Sztrajman:2021:Neural}, and our representation network. The visualization of the outgoing radiance and the difference maps are also provided. For single-layer materials, the reference is produced by the microfacet model. For layered materials, the reference images are rendered using Guo et al.~\shortcite{Guo:2018:Layered}. Both Sztrajman et al.~\shortcite{Sztrajman:2021:Neural} and our method produce similar results to the reference, while Rainer et al.~\shortcite{Rainer2020Unified} is less accurate. Note that, although the method of Sztrajman et al.~\shortcite{Sztrajman:2021:Neural} can produce smaller numerical error in some cases, it also produces obvious color differences. Also note that we focus on representation in this figure; the multi-layer BRDFs being projected are ground truth, not the outputs of our layering network.}

\subsection{Dataset}

We use the Mitsuba Renderer~\cite{Mitsuba} to generate the training dataset of BRDFs, consisting of three types: rough conductor, rough dielectric and layered BRDFs~\cite{Guo:2018:Layered}. The dataset could be enriched with other types of BRDFs as needed for the application. All BRDFs have a single channel; RGB color is achieved by independently processing each channel with varying parameters. We generate 300 rough conductor BRDFs and 300 rough dielectric BRDFs, together with $12,720$ layered BRDFs~\cite{Guo:2018:Layered} by randomly layering them into 2 layers; additionally, we also generate $1,800$ three-layer BRDFs, which means that their bottom layers are already layered BRDFs. We use the three-layer BRDFs to finetune the layering network, after it has been first trained with two layers, as described in Sec.~\ref{sec:quality}.

\begin{table}
    \centering
    \setlength{\tabcolsep}{12pt}
    \caption{Different distributions that we use to sample the parameter space of BRDFs. $\mathcal{U}(x, y)$ represents a continuous uniform distribution in the interval $(x, y)$, and $\mathcal{V}(X)$ is a discrete uniform random variable in a finite set $X$.}
    \begin{tabular}{cc} 
    \toprule
        Parameter & Sampling Function \\
      \midrule
        \makecell[c]{Roughness for rough \\ conductor/dielectric BRDFs}
        & $\alpha = \mathcal{U}(0.216, 1)^3$ \\
      
        \specialrule{0em}{2pt}{2pt}
        
        IOR for rough dielectric BSDFs
        & $\eta = \mathcal{U}(1.05, 2)$ \\

        \specialrule{0em}{2pt}{2pt}
        
        Fresnel for rough conductor BRDFs
        & $R_0 = \mathcal{U}(0, 1)$\\
        
        \specialrule{0em}{2pt}{2pt}

        Albedo for layered BRDFs
        & $A = 1 - \mathcal{U}(0, 1)^2$ \\

        \specialrule{0em}{2pt}{2pt}

        \makecell{Extinction coefficient \\ for layered BRDFs}
        & $\sigma_T = \mathcal{V}(\{0, 1, 2, 5\})$ \\
      \bottomrule
    \end{tabular}
    \label{tab:distribution}
\end{table}

Each rough dielectric BRDF has two parameters: roughness $\alpha_1$ and the index of refraction (IOR) $\eta$. Each rough conductor BRDF also has two parameters: roughness $\alpha_2$ and a Schlick Fresnel approximation with $R_0$ controlling the reflectance at $0$ degrees. Therefore, each two-layer BRDF has six parameters: $\alpha_1$, $\eta$, $\alpha_2$, $R_0$, the albedo $A$ and the extinction coefficient $\sigma_T$. In our implementation, we use GGX model as the normal distribution function at interfaces, and only consider isotropic BRDFs, although anisotropic materials could be included. The sampling distributions of different parameters are shown in Table~\ref{tab:distribution}.

For each BRDF, we sample $25^4$ pairs of incoming and outgoing directions. Since we only consider the reflection, we perform a stratified sampling along the elevation angle $\theta$ and azimuth angle $\varphi$ on the upper hemisphere, where $\theta\in[0, \frac{\pi}{2})$ and $\varphi \in [0, 2\pi)$. For each sampled incoming and outgoing direction pair, we compute the BRDF value via microfacet model for rough conductor/dielectric BRDFs or Guo et al.~\shortcite{Guo:2018:Layered} for layered BRDFs. Then we store the incoming direction, outgoing direction and the BRDF value (without the cosine term).

The generated dataset is used for training the different networks:

\paragraph{Representation network} We use $12,000$ two-layer BRDFs as our training set, and the other $720$ two-layer BRDFs as validation. Note that although we do not use rough conductor/dielectric BRDFs to train this network, they can still be represented, because their appearances could be regarded as special cases of layered BRDFs.

\paragraph{Layering network} We project $12,720$ two-layer BRDFs and their components ($300$ rough conductor and $300$ rough dielectric BRDFs) from the dataset into the latent space with the trained representation network. Then the projected latent vectors are used for training the layering network, with the same proportions of training and validating set as in the representation network. Also, the $1,800$ three-layer BRDFs are used to finetune this network.

\paragraph{Sampling network} We randomly choose $3,000$ two-layer BRDFs from the dataset for training and another $300$ two-layer BRDFs for validation. We first project them into the latent space, similar to the layering network, and then we compute the ground-truth GNDFs, as mentioned in Sec.~\ref{subsec:sample}.

\subsection{Training}
The representation network is trained first, as the layering and sampling networks rely on the representation network.

\paragraph{Representation Network} During training this network, the shared network parameters and the latent vectors of current input BRDFs are updated simultaneously in every iteration. We calculate the $L_1$ loss, which can better preserve the color and avoid artifacts, compared to the $L_2$ criterion, according to our experiments. The $L_1$ loss is simply

\begin{equation}
  Loss = \frac{1}{N}\sum_N{|f^\mathrm{pred} - f^\mathrm{gt}|},
  \label{eq:l1loss}
\end{equation} \\ where $N$ denotes the number of BRDFs in a batch, $f^{pred}$ and $f^{gt}$ represents BRDF values output by our network and the ground-truth. We use learning rates $3\times10^{-4}$ for the network weights and $1\times10^{-4}$ for the mutable latent vectors in training set. Both learning rates decay by $0.9$ after every epoch. We initialize all the latent vectors to $1$, and we train the network for $50$ epochs in about $40$ hours on an RTX 2080Ti GPU.

\paragraph{Layering Network} We supervise this network with latent vectors as both inputs and outputs, via the latent vectors obtained from our trained representation network, and optimize it by the $L_1$ loss (Equation~\ref{eq:l1loss}) with the initial learning rate of $3\times10^{-3}$. The learning rate decays by $0.7$ for every $50$ epochs. It takes us about 10 hours to train this network on an RTX 2080Ti GPU for $1,000$ epochs. Additionally, we finetune the layering network with $1,800$ three-layer BRDFs.

\paragraph{Sampling Network} We sample $40\times40$ points on the ground-truth pdf and our proxy pdf (Equation~\ref{eq:pdf}), and then minimize their difference by the Kullback-Leibler divergence (KLD) loss. The KLD loss is defined as

\begin{equation}
    KLDLoss = \frac{1}{M}\sum_{M}{\Big(\mathcal{S} (f^\mathrm{gt}) ( \textrm{log}\mathcal{S} (f^\mathrm{gt}) - \textrm{log}\mathcal{S} (f^\mathrm{pred}) )\Big)},
    \label{eq:kldloss}
\end{equation} \\ where $\mathcal{S}$ denotes the softmax function and $M$ denotes the sample count on a distribution. We start at a learning rate of $3\times10^{-5}$ and shrink it by $0.7$ for every $3$ epochs. We trained this network for $10$ epochs in total, which costs less than 1 hour on an RTX 2080Ti video card.

\mycfigure{layer_table1}{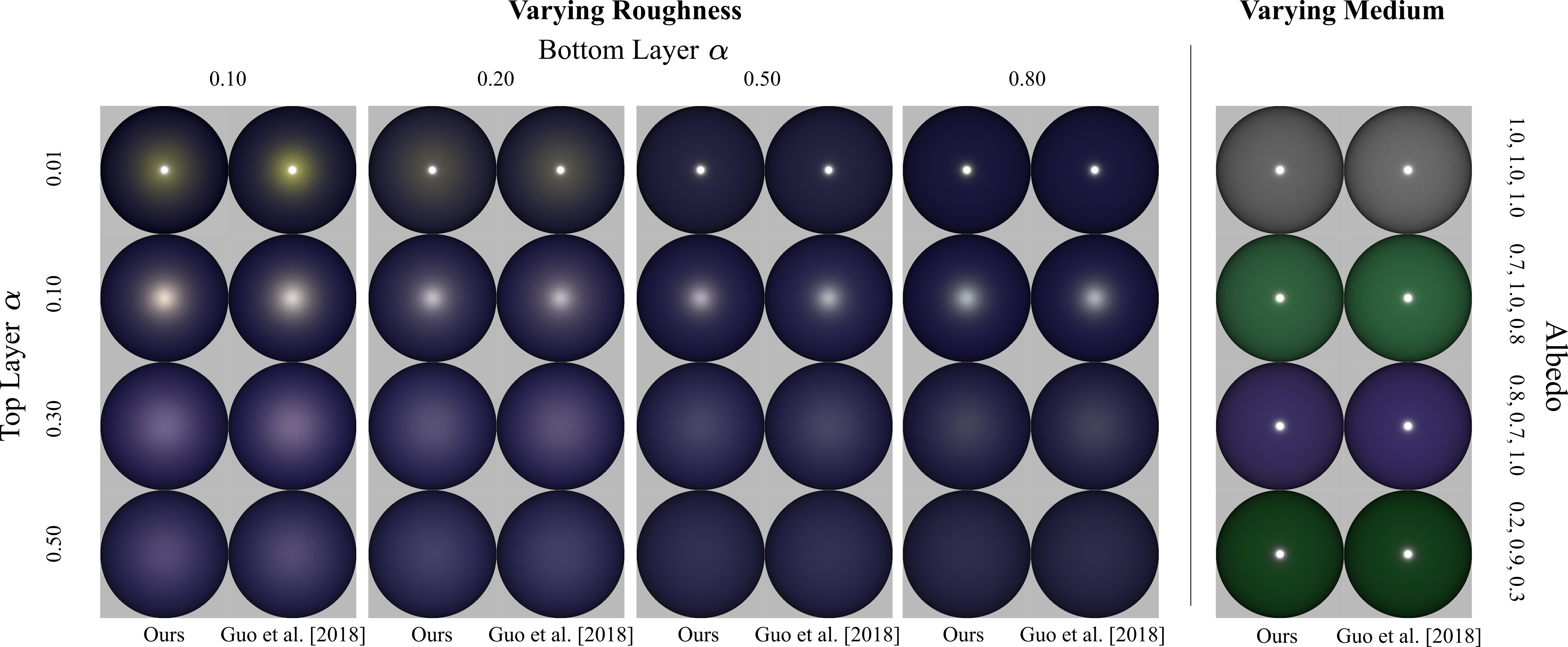}{Comparison of outgoing radiance distributions (with fixed incoming direction at zero degrees) between our layering model and Guo et al.~\shortcite{Guo:2018:Layered} on a number of BRDFs, considering varying roughness for both top and bottom layers and varying scattering albedos for the medium. Under all configurations, our layering network produces results close to the reference.}

\subsection{Rendering pipeline}

Before using our Neural BRDF models for rendering, we do some preparation: for rough conductor/dielectric BRDFs and the components of layered BRDFs, we represent them by the latent vector using our representation network; for layered BRDFs, we perform the layering operation and get the layered latent vector; for SVBRDFs, we prepare a latent texture, where each texel stores a latent vector. 

Now, we use our Neural BRDF models in rendering by path tracing. Firstly, we store all incoming and outgoing directions and lighting values of each ray at all intersections into buffers. Secondly, for each buffer pixel with the Neural BRDF type, we infer the representation network for the BRDF value. Finally, we calculate the radiance of path tracing to get output images, applying specific reconstruction filters, such as Gaussian filters. 

We also integrate our implementation with the MIS framework. According to the different sampling demands (light sampling, BRDF sampling or MIS), we store all queries and the pdf information for each part above into buffers. Then we calculate the radiance via our network and finally combine the weighted results, if needed.

In order to accelerate the GPU inference, we implement the inference in CUDA via NVIDIA CUTLASS CUDA Template and compile it into Python libraries. Thanks to the delicate optimization in Cutlass, there is no obvious increase in the time cost when the film resolution rises, as long as it doesn't run out of the GPU memory. We compile several libraries for different buffer sizes in advance, and dynamically decide which to use during rendering. Eventually, the single BRDF evaluation with resolution $1920\times1080$ costs 5 ms.
\section{Results}
\label{sec:result}

We have implemented our method inside the Mitsuba renderer \cite{Mitsuba} and compared our method with previous works, including Guo et al.~\shortcite{Guo:2018:Layered} and Belcour~\shortcite{Belcour:2018:Layered}. Specifically, since the method by Guo et al.~\shortcite{Guo:2018:Layered} does not introduce any approximations other than Monte Carlo noise, we use it as the reference. All the implementations are taken from the authors' websites. All timings in this section are measured on an Ubuntu Linux workstation with an Intel Xeon E5-2650 v4 @ 2.20GHz CPU (8 cores), 64 GB of main memory and an RTX 2080Ti GPU (11 GB).

\subsection{Quality validation}
\label{sec:quality}

\myfigure{merl}{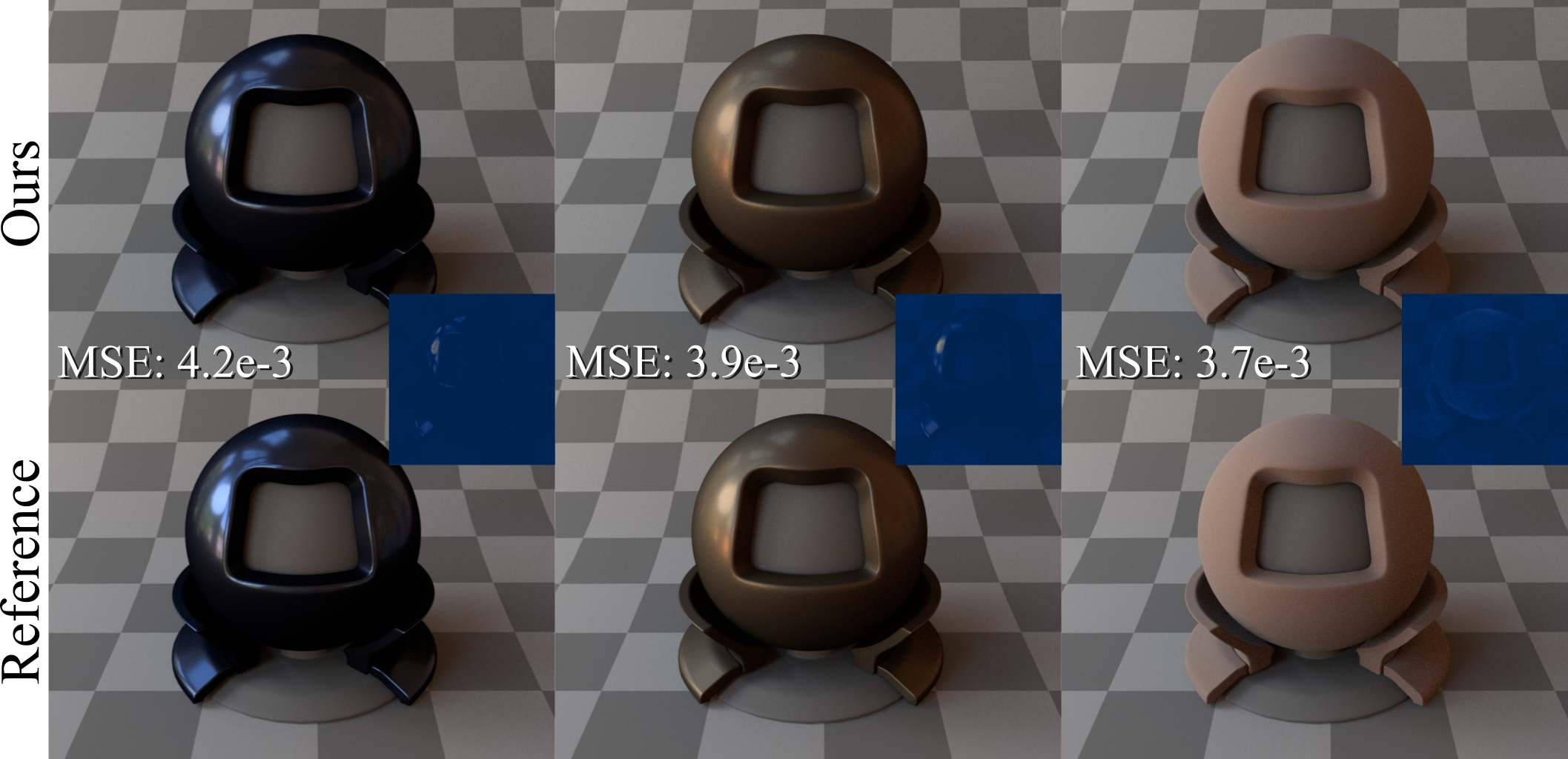}{Our method is able to represent both rough and smooth BRDFs from the MERL dataset~\cite{matusik2003data}. The reference is rendered using interpolated BRDF queries from the dataset directly.}

\myfigure{BTF}{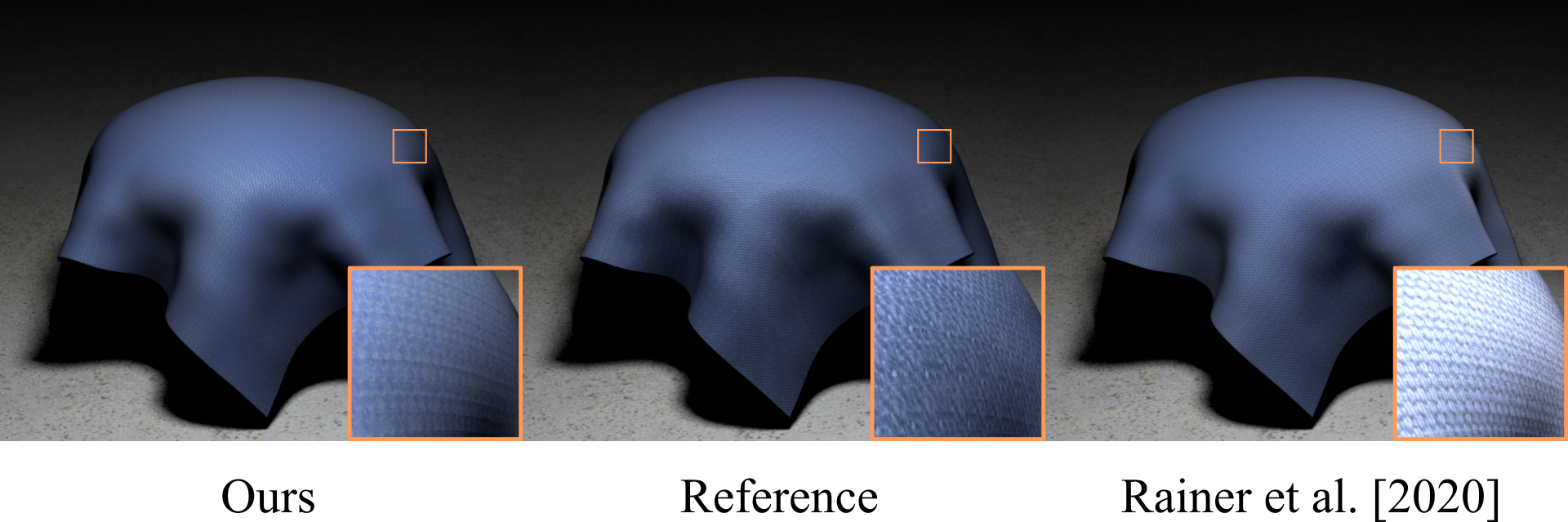}{Comparison of the representation ability on a measured SVBRDF/BTF using our method (left), Rainer et al.~\shortcite{Rainer2020Unified} (right) and the reference (Guo et al.~\shortcite{Guo:2018:Layered}, middle). The BTF data is from the UBO2014 BTF dataset~\cite{weinmann-2014-materialclassification}. While the global appearance is good using both methods, our method produces better grazing-angle result than Rainer et al.~\shortcite{Rainer2020Unified}. All insets are with $+2.8$ exposure adjustment.}

We first validate our individual operations, then demonstrate rendering results on more complex scenes.

\paragraph{Representation/evaluation network.}
In Figure~\ref{fig:rep_matpreview}, we compare our representation network against Rainer et al.~\shortcite{Rainer2020Unified}, Sztrajman et al.~\shortcite{Sztrajman:2021:Neural} and Guo et al.~\shortcite{Guo:2018:Layered} (reference) on varying materials. We use the pretrained model provided by Rainer et al.~\shortcite{Rainer2020Unified} and train the model for Sztrajman et al.~\shortcite{Sztrajman:2021:Neural} using their released code. Rainer et al.~\shortcite{Rainer2020Unified} cannot handle high-frequency materials well, and suffers from visible artifacts.

Although Sztrajman et al.~\shortcite{Sztrajman:2021:Neural} produces visually similar results to ours, their method has $6\times$ more storage cost than our method (even when using the dimensionality reduction), which makes it less practical for SVBRDFs. Even though Sztrajman et al.~\shortcite{Sztrajman:2021:Neural} can produce lower error in some cases (rows $2$--$4$), there are still obvious color differences from the reference. 

Our representation network is suitable not only for analytic BRDF data, but can generalize to measured BRDFs, such as those from the MERL dataset~\cite{matusik2003data}. In Figure~\ref{fig:merl}, we fine-tune our trained network on the MERL dataset for 30 epochs in 3 hours, and we provide the visualization of the outgoing radiance distributions from our representation network. We also show that our representation network is able to represent SVBRDFs/BTFs in Figure~\ref{fig:BTF} (note again that Sztrajman et al.~\shortcite{Sztrajman:2021:Neural} would be difficult to apply for this purpose).

\paragraph{Sampling network.}
In Figure~\ref{fig:sampling_compare}, we compare the rendered results using our sampling network against sampling parametric lobes like Lambertian lobes or GGX, with equal number of samples per pixel. Our method produces the best results. Also note that, since our fitted GNDF is independent of the incoming direction, and only dependent on the underlying BRDF, we can precompute and store the fits as a preprocessing. In this way, we avoid the expensive inference of the sampling network when drawing samples at rendering time. Therefore, our sampling method is as efficient as sampling an analytic BRDF lobe. 

Having enabled BRDF sampling, our method automatically enables MIS. In Figure~\ref{fig:MIS_validate}, we compare the results rendered with light sampling only, BRDF sampling only, and MIS combining light sampling and BRDF sampling. As expected, MIS further improves the sampling quality.

\myfigure{ball}{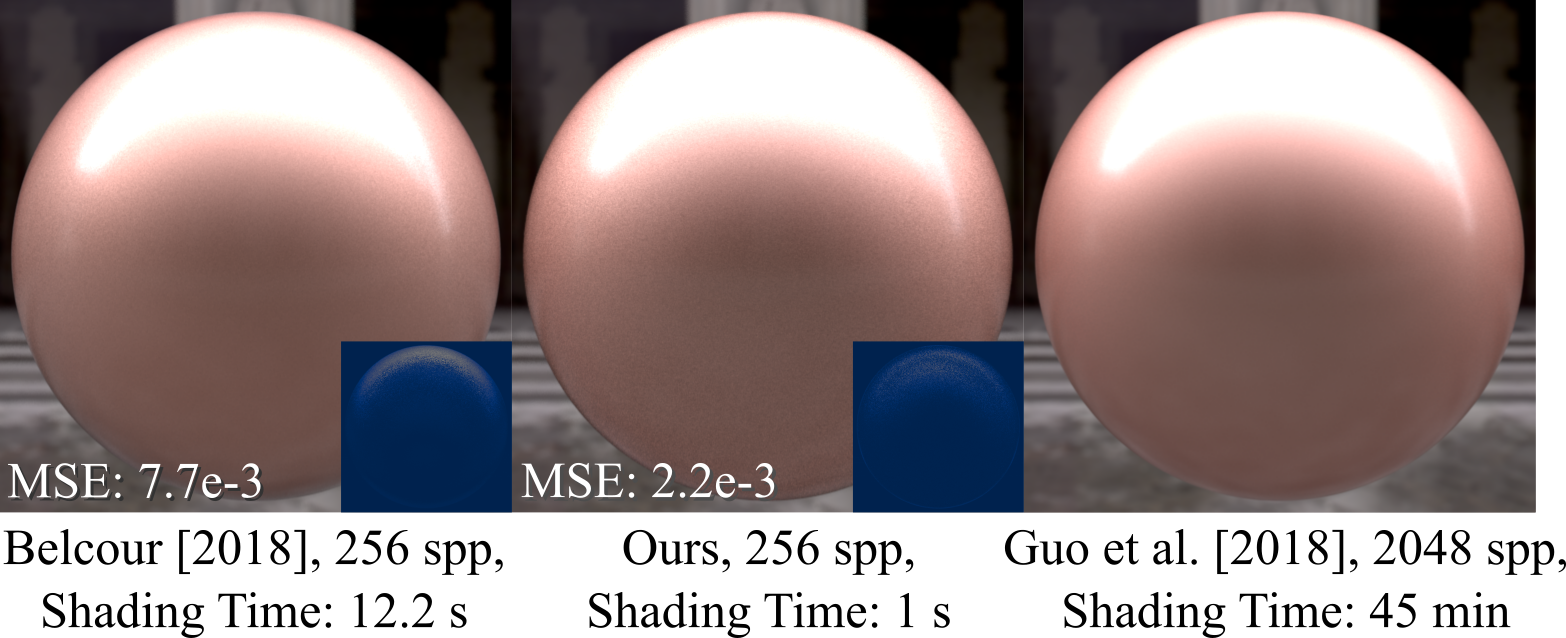}{Comparison between Belcour~\shortcite{Belcour:2018:Layered}, our method and Guo et al.~\shortcite{Guo:2018:Layered} (reference). With equal number of samples, our method produces closer result to the reference, as compared to Belcour~\shortcite{Belcour:2018:Layered}. Note that all three methods have the same cost for path tracing at the same sampling rate ($37.5$ s for $256$ spp), while our shading time is shorter (though the other methods are CPU-only while our hybrid method uses the GPU for shading).}

\myfigure{3layers}{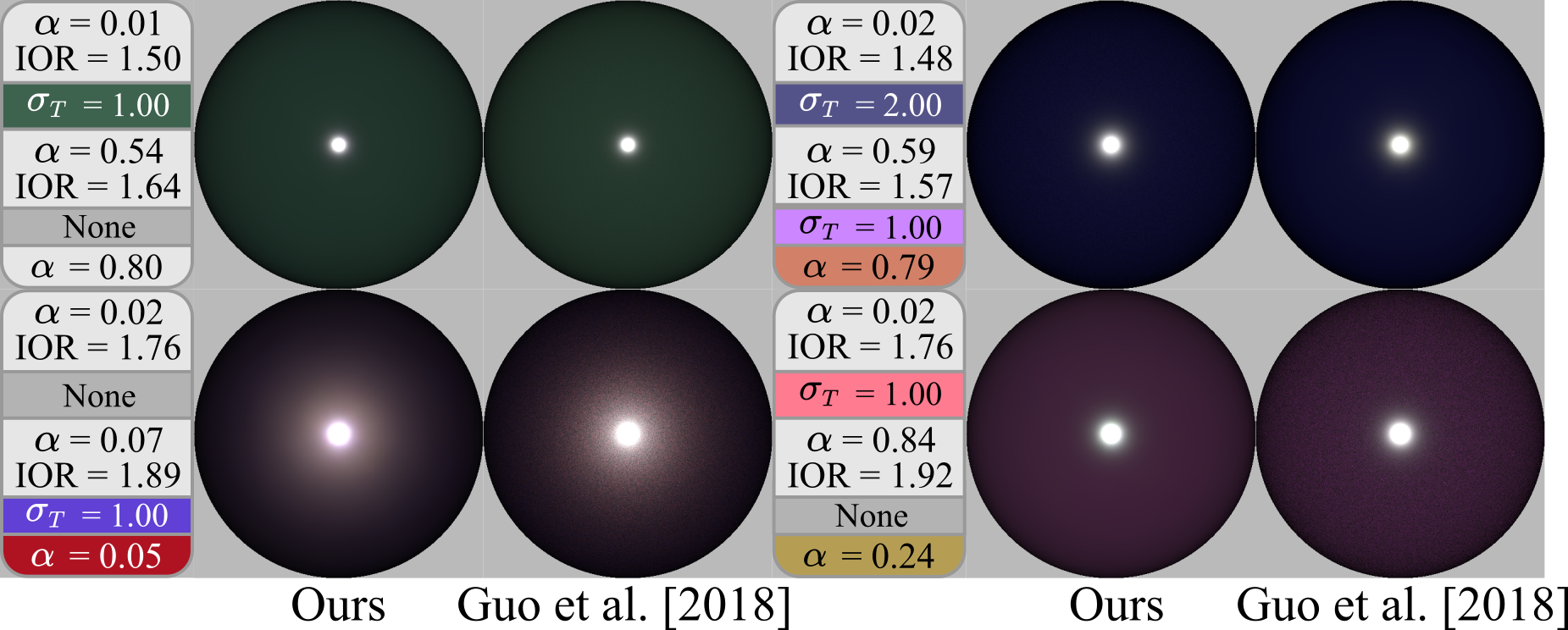}{We perform the layering recursively from bottom to top, to obtain three-layer materials. By comparing the visualization of outgoing radiance of our layering results and those from Guo et al.~\shortcite{Guo:2018:Layered}, we find the results are close to the reference.}

\paragraph{Layering network.}
In Figure~\ref{fig:layer_table1}, we compare results from our layering network against Guo et al.~\shortcite{Guo:2018:Layered} (as reference with a high sampling rate) across different BRDF configurations --- different roughness for the top and bottom layers and varying albedos for the medium in between. Our results are close to the reference. In Figure~\ref{fig:ball}, we compare our layered materials with Belcour~\shortcite{Belcour:2018:Layered} and Guo et al.~\shortcite{Guo:2018:Layered}. We can see that our method produces results close to the reference and with less noise, and our shading speed is faster than Belcour~\shortcite{Belcour:2018:Layered} (though note that we utilize the GPU for network inference). This confirms the effectiveness of our layering network.

Our layering network can be applied recursively to obtain materials with multiple (3or more) layers. In Figure~\ref{fig:3layers}, we show the results with three-layer BRDFs using our layering network fine-tuned on $1,800$ of them, and we find our result still close to the reference.

\paragraph{Interpolation.} 
Our neural BRDFs support interpolation in the latent space. We compare the lobe visualizations of our method and the linear interpolation (blending/mixture) in Figure~\ref{fig:interpolation}. Our latent space interpolation gives more natural results. For example, during the interpolation between two BRDFs with low roughness and high roughness, we naturally expect one lobe with intermediate roughness, rather than a mixture of both.

We further extend the BRDF interpolation operation to level-of-detail rendering. Recall that we use a multi-channel texture to define SVBRDFs, where each texel is a latent vector instead of an RGB/RGBA value. In Figure~\ref{fig:mipmap}, we build a mipmap of our latent texture as a preprocessing, then we use standard trilinear interpolation to query the mipmap at appropriate levels during rendering, which successfully avoids the aliasing even at a low sampling rate (1 spp). 

\subsection{Complex scenes}

\mycfigure{globe}{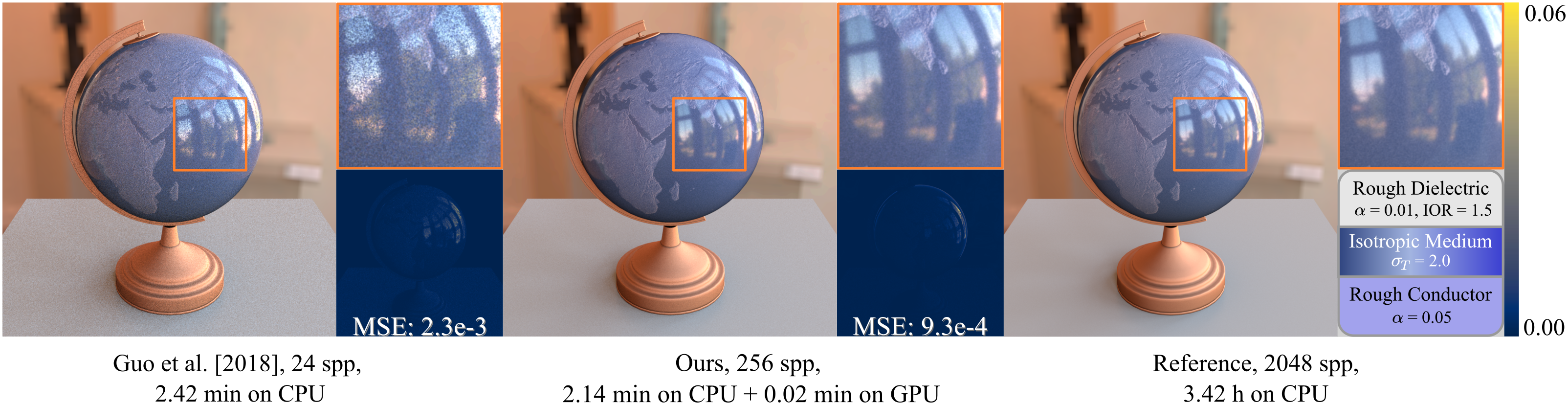}{Comparison between Guo et al.~\shortcite{Guo:2018:Layered} and our method on the Globe scene with spatially-varying albedos of the medium between two layers. The difference images show that our method has much less error than Guo et al.~\shortcite{Guo:2018:Layered}.}

\mycfigure{teapot}{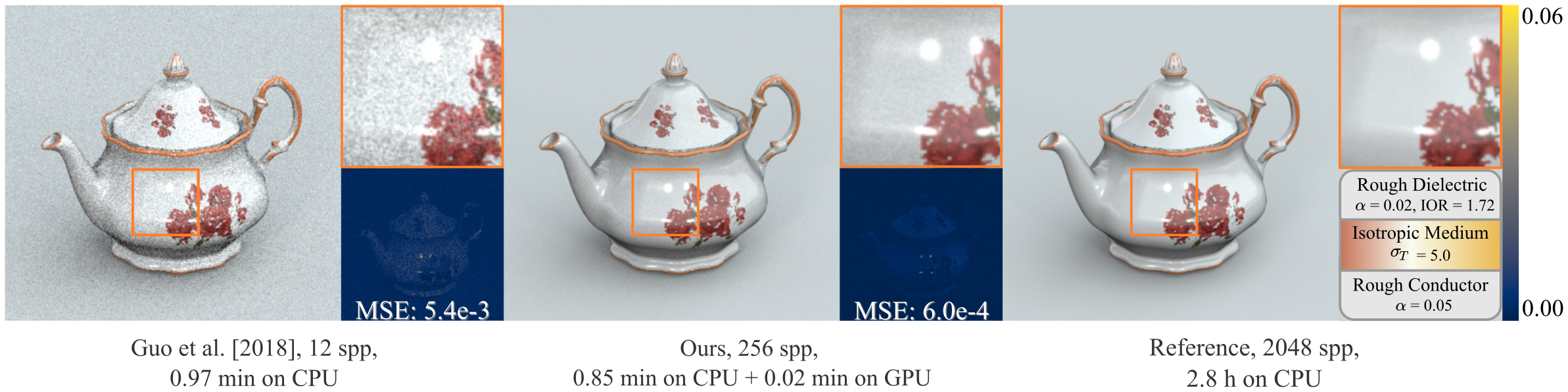}{Comparison between Guo et al.~\shortcite{Guo:2018:Layered} and our method on the Teapot scene with spatially-varying albedos of the medium between two layers (and spatially-varying roughness in the video). Again, the difference images show that our method has much less error than Guo et al.~\shortcite{Guo:2018:Layered}.}

\myfigure{shoe}{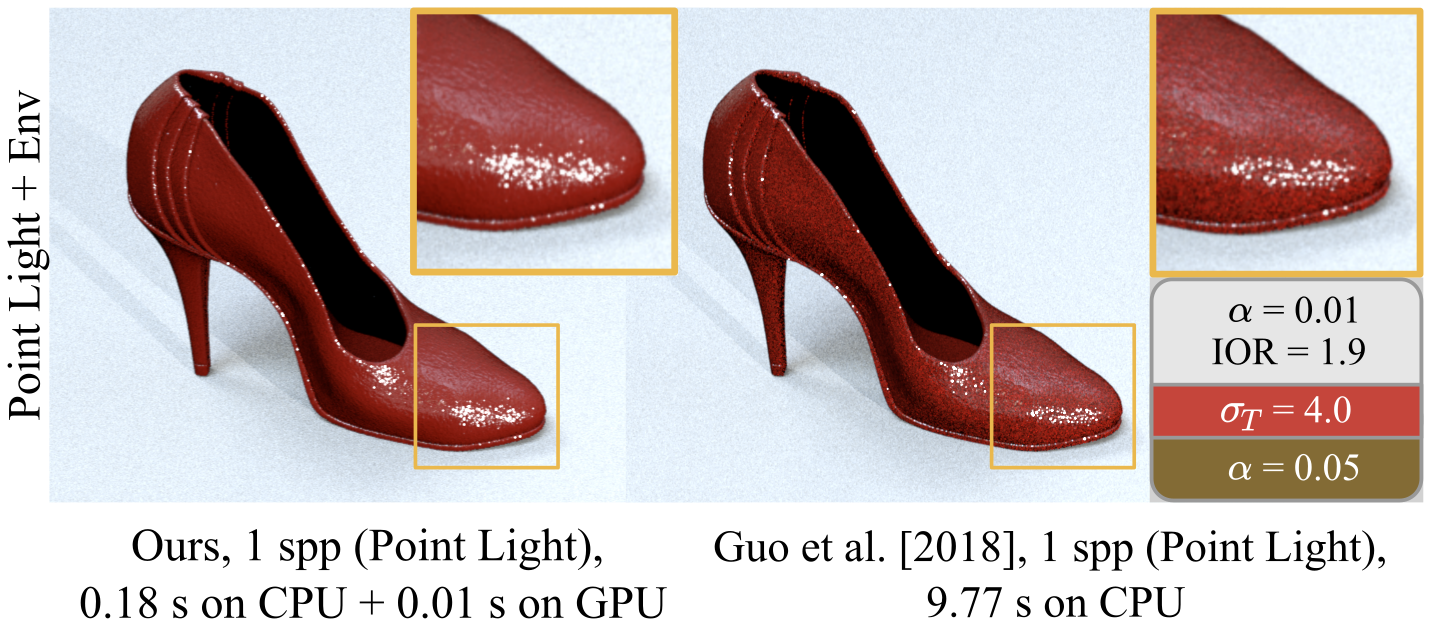}{Comparison between our method and Guo et al.~\shortcite{Guo:2018:Layered} in the Shoe scene. In this scene, we apply a normal map on the layered BRDF, and we render both methods with 1 spp. Thanks to our noise-free BRDF layering and evaluation operations, our method has low noise even at 1 spp (noise only comes from the environment lighting and indirect illumination).}

In Table~\ref{tab:performance}, we report the scene settings (and additional equal time MSE comparison with Guo et al.~\shortcite{Guo:2018:Layered}). Please also check out the accompanying video, where we show animations of the complex scenes and elaborate the detailed parameters of our layered BRDFs. Again, the cost of our method includes CPU time and GPU time, and the GPU is only used for inference of our neural networks.

\paragraph{Globe.} In Figure~\ref{fig:globe}, we show a globe with a two-layer BRDF. The top of the BRDF is a relatively smooth dielectric varnish layer and the bottom is a rough conductor. The medium between two layers has spatially-varying scattering properties. And we define spatially-varying albedos of the interface between two layers. In this scene, we compare our method against Guo et al.~\shortcite{Guo:2018:Layered} at equal time. We find that our result has much less noise, since our method does not use randomness during BRDF evaluation.

\paragraph{Teapot.} In Figure~\ref{fig:teapot}, we demonstrate the ability of our method to perform appearance editing by changing the underlying latent texture. The BRDF consists of a varnish layer on the top and a rough conductor layer on the bottom. We can arbitrarily specify spatially-varying scattering parameters (e.g., albedos) in the medium interface between layers (as well as spatially-varying roughness, shown in the accompanying video) to display various patterns. Each edit requires a re-evaluation of our layering network, which is fast compared to the rendering itself. Our method is able to produce results close to the ground-truth for the edited spatially-varying material, and our method produces much less noise than Guo et al.~\shortcite{Guo:2018:Layered}.

\paragraph{Shoe.} In Figure~\ref{fig:shoe}, we compare our method with Guo et al.~\shortcite{Guo:2018:Layered} on the Shoe scene under a point light and environment lighting. The surface of the Shoe is defined with a normal mapped BRDF consisting of two layers and a constant medium in between. With only 1 spp, our method is already close to noise-free (since the highlights mostly come from the point light in direct illumination), thanks to our noise-free layering and evaluation operations.

\paragraph{Still Life.} Figure~\ref{teaser} shows a variety of spatially-varying effects, including varying roughness, varying Fresnel, varying albedos of the interface, and normal mapping. Again, we can see that our method produces almost identical results to the reference on all these configurations within much less time.

\begin{table}
    \centering
    \setlength{\tabcolsep}{1pt}
    \caption{We tabulate the scene settings, time consumption (minutes) and error (MSE) in all scenes, compared with Guo et al.~\shortcite{Guo:2018:Layered}.}
    \label{tab:performance}
    \begin{tabular}{ccccccc}
        \toprule 
        \multirow{2}{*}[-0.5ex]{Scene}  & \multirow{2}{*}[-0.5ex]{Resolution}   & \multicolumn{2}{c}{Ours}  & \multicolumn{2}{c}{Guo et al.\shortcite{Guo:2018:Layered}}  & \multirow{2}{*}[-0.5ex]{Time}   \\
                                                                                \cmidrule(l{1pt}r{1pt}){3-4}   \cmidrule(l{1pt}r{1pt}){5-6}
                                        &                                       &   Spp  & MSE                & Spp   &   MSE \\
        \midrule
        Still Life (Fig.~\ref{teaser}) & $1024\times1024$ & 512 & \bm{$1.0\times10^{-3}$} & 64 & $2.0\times10^{-3}$ & 4.56 \\
        Globe   (Fig.~\ref{fig:globe}) & $1024\times1024$ & 256 & \bm{$9.3\times10^{-4}$} & 24 & $2.3\times10^{-3}$ & 2.16 \\ 
        Teapot   (Fig.~\ref{fig:teapot}) & $720\times480$ & 256 & \bm{$6.0\times10^{-4}$} & 12 & $5.4\times10^{-3}$ & 0.87 \\ 
        \bottomrule
    \end{tabular}
\end{table}

\subsection{Discussion and limitations}
\label{sec:discussion}
Our method is able to represent a large range of single-layer and mutli-layer BRDFs with both specular and diffuse appearances, and the latent representation can be easily operated to produce different effects. However, there are some approaches that we have not tried but may be potentially helpful, and we also have identified some main limitations of our method. Below, we briefly discuss the key points.

\paragraph{Bidirectional Transmittance Distribution Functions (BTDFs)} When training our layering network, we only use the reflection information from each BRDF layer to get the reflectance of the final layered material. When a BRDF is used as the top interface of a layering, its corresponding BTDF will be implicitly inferred by the layering network, and is never constructed explicitly. On the one hand, this is an advantage of our method, because we will never need to store and represent BTDFs. On the other hand, an extension to full ``Neural BSDFs'' may be interesting and useful for other effects (e.g., translucent fabrics).

\paragraph{Scope/types of BRDFs} Currently, we do not train our model on anisotropic or normal mapped BRDFs (from individual layers). Including these types of BRDFs will further improve the practicality of our method, but can also be more challenging to both neural representation and operation, since the dimension of input data will further increase. While we believe our general framework could handle these effects, we leave the exploration of anisotropic and normal mapped BRDFs for future work.

\emph{Accumulated error}. We have shown that when recursively applied, our layering network can be used to predict BRDFs consisted of multiple layers. However, error will accumulate as more BRDFs are layered together. This could be addressed by networks trained on a larger fixed number of BRDF layers, or exploiting neural network architectures that allow a dynamic number of inputs.

\emph{Bias and energy conservation}. In practice, any neural network may produce unpredictable error, which cannot be treated as unbiased in the Monte Carlo sense. The error can make the results consistently darker/brighter in some angular regions, introducing bias or violation of energy conservation in rendering. This issue affects all neural rendering solutions; we have not observed artifacts caused by this, but the problem may require future research.

\section{Conclusion and Future Work}
\label{sec:conclusion}

In this paper, we have presented a framework for neural BRDFs. We focus on both representation and operations for BRDFs. We use a general neural network to compress BRDFs into short latent vectors, and we train additional networks that operate solely in the latent space, providing individual operations to BRDFs. Our representation network is able to compress a wide range of BRDFs, including typical microfacet BRDFs, layered BRDFs or measured BRDFs, accurately and compactly. And we have demonstrated those operation networks are able to perform common BRDF operations, such as importance sampling, interpolation and layering. Eventually, our Neural BRDFs can be easily used in the rendering pipeline under the MIS framework. And we have shown that shading using our BRDF latent textures and operation neural networks is efficient, especially for layered materials.

We believe that the proposed representation/operations model is novel and practical, leading to a black-box style ``Neural BRDF algebra''. However, this BRDF algebra is still not complete enough to encompass all common BRDFs and operations on them. Our model is currently trained on isotropic materials and media for now, but it could be extended to anisotropic materials and media, as well as explicit handling of transmissive BTDFs. It would also be interesting to introduce normal mapping operations in latent space.

\bibliographystyle{ACM-Reference-Format}
\bibliography{paper}


\begin{thebibliography}{36}


\ifx \showCODEN    \undefined \def \showCODEN     #1{\unskip}     \fi
\ifx \showDOI      \undefined \def \showDOI       #1{#1}\fi
\ifx \showISBNx    \undefined \def \showISBNx     #1{\unskip}     \fi
\ifx \showISBNxiii \undefined \def \showISBNxiii  #1{\unskip}     \fi
\ifx \showISSN     \undefined \def \showISSN      #1{\unskip}     \fi
\ifx \showLCCN     \undefined \def \showLCCN      #1{\unskip}     \fi
\ifx \shownote     \undefined \def \shownote      #1{#1}          \fi
\ifx \showarticletitle \undefined \def \showarticletitle #1{#1}   \fi
\ifx \showURL      \undefined \def \showURL       {\relax}        \fi
\providecommand\bibfield[2]{#2}
\providecommand\bibinfo[2]{#2}
\providecommand\natexlab[1]{#1}
\providecommand\showeprint[2][]{arXiv:#2}

\bibitem[\protect\citeauthoryear{Belcour}{Belcour}{2018}]%
        {Belcour:2018:Layered}
\bibfield{author}{\bibinfo{person}{Laurent Belcour}.}
  \bibinfo{year}{2018}\natexlab{}.
\newblock \showarticletitle{Efficient Rendering of Layered Materials Using an
  Atomic Decomposition with Statistical Operators}.
\newblock \bibinfo{journal}{\emph{ACM Trans. Graph.}} \bibinfo{volume}{37},
  \bibinfo{number}{4}, Article \bibinfo{articleno}{73} (\bibinfo{date}{July}
  \bibinfo{year}{2018}), \bibinfo{numpages}{15}~pages.
\newblock


\bibitem[\protect\citeauthoryear{Bi, Xu, Srinivasan, Mildenhall, Sunkavalli,
  Hašan, Hold-Geoffroy, Kriegman, and Ramamoorthi}{Bi et~al\mbox{.}}{2020}]%
        {Bi:2020:Appearance}
\bibfield{author}{\bibinfo{person}{Sai Bi}, \bibinfo{person}{Zexiang Xu},
  \bibinfo{person}{Pratul Srinivasan}, \bibinfo{person}{Ben Mildenhall},
  \bibinfo{person}{Kalyan Sunkavalli}, \bibinfo{person}{Miloš Hašan},
  \bibinfo{person}{Yannick Hold-Geoffroy}, \bibinfo{person}{David Kriegman},
  {and} \bibinfo{person}{Ravi Ramamoorthi}.} \bibinfo{year}{2020}\natexlab{}.
\newblock \bibinfo{title}{Neural Reflectance Fields for Appearance
  Acquisition}.
\newblock
\newblock


\bibitem[\protect\citeauthoryear{Dana, van Ginneken, Nayar, and
  Koenderink}{Dana et~al\mbox{.}}{1999}]%
        {Dana:1999:BTF}
\bibfield{author}{\bibinfo{person}{Kristin~J. Dana}, \bibinfo{person}{Bram van
  Ginneken}, \bibinfo{person}{Shree~K. Nayar}, {and} \bibinfo{person}{Jan~J.
  Koenderink}.} \bibinfo{year}{1999}\natexlab{}.
\newblock \bibinfo{journal}{\emph{ACM Trans. Graph.}} \bibinfo{volume}{18},
  \bibinfo{number}{1} (\bibinfo{date}{Jan.} \bibinfo{year}{1999}),
  \bibinfo{pages}{1–34}.
\newblock


\bibitem[\protect\citeauthoryear{Gamboa, Gruson, and Nowrouzezahrai}{Gamboa
  et~al\mbox{.}}{2020}]%
        {Gamboa:2020:EfficientLayered}
\bibfield{author}{\bibinfo{person}{Luis~E. Gamboa}, \bibinfo{person}{Adrien
  Gruson}, {and} \bibinfo{person}{Derek Nowrouzezahrai}.}
  \bibinfo{year}{2020}\natexlab{}.
\newblock \showarticletitle{{An Efficient Transport Estimator for Complex
  Layered Materials}}.
\newblock \bibinfo{journal}{\emph{Computer Graphics Forum}}
  \bibinfo{volume}{39}, \bibinfo{number}{2} (\bibinfo{year}{2020}),
  \bibinfo{pages}{363--371}.
\newblock
\showISSN{1467-8659}
\urldef\tempurl%
\url{https://doi.org/10.1111/cgf.13936}
\showDOI{\tempurl}


\bibitem[\protect\citeauthoryear{Ge, Wang, Wang, Meng, and Holzschuch}{Ge
  et~al\mbox{.}}{2021}]%
        {Ge:2021:multiple}
\bibfield{author}{\bibinfo{person}{Liangsheng Ge}, \bibinfo{person}{Beibei
  Wang}, \bibinfo{person}{Lu Wang}, \bibinfo{person}{Xiangxu Meng}, {and}
  \bibinfo{person}{Nicolas Holzschuch}.} \bibinfo{year}{2021}\natexlab{}.
\newblock \showarticletitle{Interactive Simulation of Scattering Effects in
  Participating Media Using a Neural Network Model}.
\newblock \bibinfo{journal}{\emph{IEEE Transactions on Visualization and
  Computer Graphics}} \bibinfo{volume}{27}, \bibinfo{number}{7}
  (\bibinfo{year}{2021}), \bibinfo{pages}{3123--3134}.
\newblock
\urldef\tempurl%
\url{https://doi.org/10.1109/TVCG.2019.2963015}
\showDOI{\tempurl}


\bibitem[\protect\citeauthoryear{Granskog, Schnabel, Rousselle, and
  Nov{\'a}k}{Granskog et~al\mbox{.}}{2021}]%
        {granskog2021neural}
\bibfield{author}{\bibinfo{person}{Jonathan Granskog}, \bibinfo{person}{Till~N
  Schnabel}, \bibinfo{person}{Fabrice Rousselle}, {and} \bibinfo{person}{Jan
  Nov{\'a}k}.} \bibinfo{year}{2021}\natexlab{}.
\newblock \showarticletitle{Neural scene graph rendering}.
\newblock \bibinfo{journal}{\emph{ACM Transactions on Graphics (TOG)}}
  \bibinfo{volume}{40}, \bibinfo{number}{4} (\bibinfo{year}{2021}),
  \bibinfo{pages}{1--11}.
\newblock


\bibitem[\protect\citeauthoryear{Guo, Ha\v{s}an, and Zhao}{Guo
  et~al\mbox{.}}{2018}]%
        {Guo:2018:Layered}
\bibfield{author}{\bibinfo{person}{Yu Guo}, \bibinfo{person}{Milo\v{s}
  Ha\v{s}an}, {and} \bibinfo{person}{Shuang Zhao}.}
  \bibinfo{year}{2018}\natexlab{}.
\newblock \showarticletitle{Position-Free Monte Carlo Simulation for Arbitrary
  Layered BSDFs}.
\newblock \bibinfo{journal}{\emph{ACM Trans. Graph.}} \bibinfo{volume}{37},
  \bibinfo{number}{6}, Article \bibinfo{articleno}{279} (\bibinfo{date}{Dec.}
  \bibinfo{year}{2018}), \bibinfo{numpages}{14}~pages.
\newblock


\bibitem[\protect\citeauthoryear{Jakob}{Jakob}{2010}]%
        {Mitsuba}
\bibfield{author}{\bibinfo{person}{Wenzel Jakob}.}
  \bibinfo{year}{2010}\natexlab{}.
\newblock \bibinfo{title}{Mitsuba renderer}.
\newblock
\newblock
\newblock
\shownote{http://www.mitsuba-renderer.org.}


\bibitem[\protect\citeauthoryear{Jakob}{Jakob}{2015}]%
        {Jakob2015Layerlab}
\bibfield{author}{\bibinfo{person}{Wenzel Jakob}.}
  \bibinfo{year}{2015}\natexlab{}.
\newblock \showarticletitle{layerlab: A computational toolbox for layered
  materials}. In \bibinfo{booktitle}{\emph{SIGGRAPH 2015 Courses}}
  \emph{(\bibinfo{series}{SIGGRAPH '15})}. \bibinfo{publisher}{ACM},
  \bibinfo{address}{New York, NY, USA}.
\newblock
\urldef\tempurl%
\url{https://doi.org/10.1145/2776880.2787670}
\showDOI{\tempurl}


\bibitem[\protect\citeauthoryear{Jakob, d'Eon, Jakob, and Marschner}{Jakob
  et~al\mbox{.}}{2014}]%
        {jakob:2014:layered}
\bibfield{author}{\bibinfo{person}{Wenzel Jakob}, \bibinfo{person}{Eugene
  d'Eon}, \bibinfo{person}{Otto Jakob}, {and} \bibinfo{person}{Steve
  Marschner}.} \bibinfo{year}{2014}\natexlab{}.
\newblock \showarticletitle{A Comprehensive Framework for Rendering Layered
  Materials}.
\newblock \bibinfo{journal}{\emph{ACM Trans. Graph.}} \bibinfo{volume}{33},
  \bibinfo{number}{4}, Article \bibinfo{articleno}{118} (\bibinfo{date}{July}
  \bibinfo{year}{2014}), \bibinfo{numpages}{14}~pages.
\newblock


\bibitem[\protect\citeauthoryear{Kim, Choi, and Lee}{Kim et~al\mbox{.}}{2018}]%
        {KIM2018:Compress}
\bibfield{author}{\bibinfo{person}{Yong~Hwi Kim}, \bibinfo{person}{Junho Choi},
  {and} \bibinfo{person}{Kwan~H. Lee}.} \bibinfo{year}{2018}\natexlab{}.
\newblock \showarticletitle{An efficient method for specular-enhanced BTF
  compression}.
\newblock \bibinfo{journal}{\emph{Computers \& Graphics}}  \bibinfo{volume}{75}
  (\bibinfo{year}{2018}), \bibinfo{pages}{1--10}.
\newblock


\bibitem[\protect\citeauthoryear{Koudelka, Magda, Belhumeur, and
  Kriegman}{Koudelka et~al\mbox{.}}{2003}]%
        {Koudelka03compression}
\bibfield{author}{\bibinfo{person}{Melissa~L. Koudelka},
  \bibinfo{person}{Sebastian Magda}, \bibinfo{person}{Peter~N. Belhumeur},
  {and} \bibinfo{person}{David~J. Kriegman}.} \bibinfo{year}{2003}\natexlab{}.
\newblock \showarticletitle{Acquisition, compression, and synthesis of
  bidirectional texture functions}. In \bibinfo{booktitle}{\emph{In ICCV 03
  Workshop on Texture Analysis and Synthesis}}.
\newblock


\bibitem[\protect\citeauthoryear{Kuznetsov, Ha\v{s}an, Xu, Yan, Walter,
  Kalantari, Marschner, and Ramamoorthi}{Kuznetsov et~al\mbox{.}}{2019}]%
        {Kuznetsov19:GlintsGAN}
\bibfield{author}{\bibinfo{person}{Alexandr Kuznetsov},
  \bibinfo{person}{Milo\v{s} Ha\v{s}an}, \bibinfo{person}{Zexiang Xu},
  \bibinfo{person}{Ling-Qi Yan}, \bibinfo{person}{Bruce Walter},
  \bibinfo{person}{Nima~Khademi Kalantari}, \bibinfo{person}{Steve Marschner},
  {and} \bibinfo{person}{Ravi Ramamoorthi}.} \bibinfo{year}{2019}\natexlab{}.
\newblock \showarticletitle{Learning Generative Models for Rendering Specular
  Microgeometry}.
\newblock \bibinfo{journal}{\emph{ACM Trans. Graph.}} \bibinfo{volume}{38},
  \bibinfo{number}{6}, Article \bibinfo{articleno}{225} (\bibinfo{date}{Nov.}
  \bibinfo{year}{2019}), \bibinfo{numpages}{14}~pages.
\newblock
\showISSN{0730-0301}
\urldef\tempurl%
\url{https://doi.org/10.1145/3355089.3356525}
\showDOI{\tempurl}


\bibitem[\protect\citeauthoryear{Kuznetsov, Mullia, Xu, Ha\v{s}an, and
  Ramamoorthi}{Kuznetsov et~al\mbox{.}}{2021}]%
        {kuznetsov2021neumip}
\bibfield{author}{\bibinfo{person}{Alexandr Kuznetsov},
  \bibinfo{person}{Krishna Mullia}, \bibinfo{person}{Zexiang Xu},
  \bibinfo{person}{Milo\v{s} Ha\v{s}an}, {and} \bibinfo{person}{Ravi
  Ramamoorthi}.} \bibinfo{year}{2021}\natexlab{}.
\newblock \showarticletitle{NeuMIP: Multi-Resolution Neural Materials}.
\newblock \bibinfo{journal}{\emph{Transactions on Graphics (Proceedings of
  SIGGRAPH)}} \bibinfo{volume}{40}, \bibinfo{number}{4}, Article
  \bibinfo{articleno}{175} (\bibinfo{date}{July} \bibinfo{year}{2021}),
  \bibinfo{numpages}{13}~pages.
\newblock


\bibitem[\protect\citeauthoryear{Matusik}{Matusik}{2003}]%
        {matusik2003data}
\bibfield{author}{\bibinfo{person}{Wojciech Matusik}.}
  \bibinfo{year}{2003}\natexlab{}.
\newblock \emph{\bibinfo{title}{A data-driven reflectance model}}.
\newblock \bibinfo{thesistype}{Ph.D. Dissertation}.
  \bibinfo{school}{Massachusetts Institute of Technology}.
\newblock


\bibitem[\protect\citeauthoryear{Mildenhall, Srinivasan, Tancik, Barron,
  Ramamoorthi, and Ng}{Mildenhall et~al\mbox{.}}{2020}]%
        {Mildenhall2020NeRFRS}
\bibfield{author}{\bibinfo{person}{B. Mildenhall}, \bibinfo{person}{Pratul~P.
  Srinivasan}, \bibinfo{person}{Matthew Tancik}, \bibinfo{person}{J.~T.
  Barron}, \bibinfo{person}{R. Ramamoorthi}, {and} \bibinfo{person}{Ren Ng}.}
  \bibinfo{year}{2020}\natexlab{}.
\newblock \showarticletitle{NeRF: Representing Scenes as Neural Radiance Fields
  for View Synthesis}. In \bibinfo{booktitle}{\emph{ECCV}}.
\newblock


\bibitem[\protect\citeauthoryear{M\"{u}ller, Mcwilliams, Rousselle, Gross, and
  Nov\'{a}k}{M\"{u}ller et~al\mbox{.}}{2019}]%
        {Muller:2019:sampling}
\bibfield{author}{\bibinfo{person}{Thomas M\"{u}ller}, \bibinfo{person}{Brian
  Mcwilliams}, \bibinfo{person}{Fabrice Rousselle}, \bibinfo{person}{Markus
  Gross}, {and} \bibinfo{person}{Jan Nov\'{a}k}.}
  \bibinfo{year}{2019}\natexlab{}.
\newblock \showarticletitle{Neural Importance Sampling}.
\newblock \bibinfo{journal}{\emph{ACM Trans. Graph.}} \bibinfo{volume}{38},
  \bibinfo{number}{5}, Article \bibinfo{articleno}{145} (\bibinfo{date}{Oct.}
  \bibinfo{year}{2019}), \bibinfo{numpages}{19}~pages.
\newblock


\bibitem[\protect\citeauthoryear{Nalbach, Arabadzhiyska, Mehta, Seidel, and
  Ritschel}{Nalbach et~al\mbox{.}}{2017}]%
        {Nalbach:2017:DeepShading}
\bibfield{author}{\bibinfo{person}{O. Nalbach}, \bibinfo{person}{E.
  Arabadzhiyska}, \bibinfo{person}{D. Mehta}, \bibinfo{person}{H.-P. Seidel},
  {and} \bibinfo{person}{T. Ritschel}.} \bibinfo{year}{2017}\natexlab{}.
\newblock \showarticletitle{Deep Shading: Convolutional Neural Networks for
  Screen Space Shading}.
\newblock \bibinfo{journal}{\emph{Comput. Graph. Forum}} \bibinfo{volume}{36},
  \bibinfo{number}{4} (\bibinfo{date}{July} \bibinfo{year}{2017}),
  \bibinfo{pages}{65–78}.
\newblock


\bibitem[\protect\citeauthoryear{Rainer, Ghosh, Jakob, and Weyrich}{Rainer
  et~al\mbox{.}}{2020}]%
        {Rainer2020Unified}
\bibfield{author}{\bibinfo{person}{Gilles Rainer}, \bibinfo{person}{Abhijeet
  Ghosh}, \bibinfo{person}{Wenzel Jakob}, {and} \bibinfo{person}{Tim Weyrich}.}
  \bibinfo{year}{2020}\natexlab{}.
\newblock \showarticletitle{Unified Neural Encoding of BTFs}.
\newblock \bibinfo{journal}{\emph{Computer Graphics Forum (Proceedings of
  Eurographics)}} \bibinfo{volume}{39}, \bibinfo{number}{2}
  (\bibinfo{date}{June} \bibinfo{year}{2020}).
\newblock
\urldef\tempurl%
\url{https://doi.org/10.1111/cgf.13921}
\showDOI{\tempurl}


\bibitem[\protect\citeauthoryear{Rainer, Jakob, Ghosh, and Weyrich}{Rainer
  et~al\mbox{.}}{2019}]%
        {Rainer2019Neural}
\bibfield{author}{\bibinfo{person}{Gilles Rainer}, \bibinfo{person}{Wenzel
  Jakob}, \bibinfo{person}{Abhijeet Ghosh}, {and} \bibinfo{person}{Tim
  Weyrich}.} \bibinfo{year}{2019}\natexlab{}.
\newblock \showarticletitle{Neural BTF Compression and Interpolation}.
\newblock \bibinfo{journal}{\emph{Computer Graphics Forum (Proceedings of
  Eurographics)}} \bibinfo{volume}{38}, \bibinfo{number}{2}
  (\bibinfo{date}{March} \bibinfo{year}{2019}).
\newblock


\bibitem[\protect\citeauthoryear{Ruiters and Klein}{Ruiters and Klein}{2009}]%
        {ruiters-2009-ksvd}
\bibfield{author}{\bibinfo{person}{Roland Ruiters} {and}
  \bibinfo{person}{Reinhard Klein}.} \bibinfo{year}{2009}\natexlab{}.
\newblock \showarticletitle{BTF Compression via Sparse Tensor Decomposition}.
\newblock \bibinfo{journal}{\emph{Computer Graphics Forum (Proc. of EGSR)}}
  \bibinfo{volume}{28}, \bibinfo{number}{4} (\bibinfo{date}{July}
  \bibinfo{year}{2009}), \bibinfo{pages}{1181--1188}.
\newblock


\bibitem[\protect\citeauthoryear{Sattler, Sarlette, and Klein}{Sattler
  et~al\mbox{.}}{2003}]%
        {Sattler:2003:Cloth}
\bibfield{author}{\bibinfo{person}{Mirko Sattler}, \bibinfo{person}{Ralf
  Sarlette}, {and} \bibinfo{person}{Reinhard Klein}.}
  \bibinfo{year}{2003}\natexlab{}.
\newblock \showarticletitle{Efficient and Realistic Visualization of Cloth.}
  \bibinfo{pages}{167--178}.
\newblock


\bibitem[\protect\citeauthoryear{Sztrajman, Rainer, Ritschel, and
  Weyrich}{Sztrajman et~al\mbox{.}}{2021}]%
        {Sztrajman:2021:Neural}
\bibfield{author}{\bibinfo{person}{Alejandro Sztrajman},
  \bibinfo{person}{Gilles Rainer}, \bibinfo{person}{Tobias Ritschel}, {and}
  \bibinfo{person}{Tim Weyrich}.} \bibinfo{year}{2021}\natexlab{}.
\newblock \showarticletitle{Neural BRDF Representation and Importance
  Sampling}.
\newblock \bibinfo{journal}{\emph{Computer Graphics Forum}}
  \bibinfo{volume}{n/a}, \bibinfo{number}{n/a} (\bibinfo{year}{2021}).
\newblock
\urldef\tempurl%
\url{https://doi.org/10.1111/cgf.14335}
\showDOI{\tempurl}


\bibitem[\protect\citeauthoryear{Takikawa, Litalien, Yin, Kreis, Loop,
  Nowrouzezahrai, Jacobson, McGuire, and Fidler}{Takikawa
  et~al\mbox{.}}{2021}]%
        {takikawa2021neural}
\bibfield{author}{\bibinfo{person}{Towaki Takikawa}, \bibinfo{person}{Joey
  Litalien}, \bibinfo{person}{Kangxue Yin}, \bibinfo{person}{Karsten Kreis},
  \bibinfo{person}{Charles Loop}, \bibinfo{person}{Derek Nowrouzezahrai},
  \bibinfo{person}{Alec Jacobson}, \bibinfo{person}{Morgan McGuire}, {and}
  \bibinfo{person}{Sanja Fidler}.} \bibinfo{year}{2021}\natexlab{}.
\newblock \showarticletitle{Neural geometric level of detail: Real-time
  rendering with implicit 3D shapes}. In \bibinfo{booktitle}{\emph{Proceedings
  of the IEEE/CVF Conference on Computer Vision and Pattern Recognition}}.
  \bibinfo{pages}{11358--11367}.
\newblock


\bibitem[\protect\citeauthoryear{Thies, Zollh\"{o}fer, and Nie\ss{}ner}{Thies
  et~al\mbox{.}}{2019}]%
        {Thies:2019:Neural}
\bibfield{author}{\bibinfo{person}{Justus Thies}, \bibinfo{person}{Michael
  Zollh\"{o}fer}, {and} \bibinfo{person}{Matthias Nie\ss{}ner}.}
  \bibinfo{year}{2019}\natexlab{}.
\newblock \showarticletitle{Deferred Neural Rendering: Image Synthesis Using
  Neural Textures}.
\newblock \bibinfo{journal}{\emph{ACM Trans. Graph.}} \bibinfo{volume}{38},
  \bibinfo{number}{4}, Article \bibinfo{articleno}{66} (\bibinfo{date}{July}
  \bibinfo{year}{2019}), \bibinfo{numpages}{12}~pages.
\newblock


\bibitem[\protect\citeauthoryear{Walter, Marschner, Li, and Torrance}{Walter
  et~al\mbox{.}}{2007}]%
        {Walter07}
\bibfield{author}{\bibinfo{person}{Bruce Walter}, \bibinfo{person}{Stephen~R.
  Marschner}, \bibinfo{person}{Hongsong Li}, {and} \bibinfo{person}{Kenneth~E.
  Torrance}.} \bibinfo{year}{2007}\natexlab{}.
\newblock \showarticletitle{Microfacet Models for Refraction Through Rough
  Surfaces} \emph{(\bibinfo{series}{EGSR 07})}. \bibinfo{pages}{195--206}.
\newblock


\bibitem[\protect\citeauthoryear{Wang, Wu, Shi, Yu, and Ahuja}{Wang
  et~al\mbox{.}}{2005}]%
        {Wang:2005:Tensor}
\bibfield{author}{\bibinfo{person}{Hongcheng Wang}, \bibinfo{person}{Qing Wu},
  \bibinfo{person}{Lin Shi}, \bibinfo{person}{Yizhou Yu}, {and}
  \bibinfo{person}{Narendra Ahuja}.} \bibinfo{year}{2005}\natexlab{}.
\newblock \showarticletitle{Out-of-Core Tensor Approximation of
  Multi-Dimensional Matrices of Visual Data}.
\newblock \bibinfo{journal}{\emph{ACM Trans. Graph.}} \bibinfo{volume}{24},
  \bibinfo{number}{3} (\bibinfo{date}{July} \bibinfo{year}{2005}),
  \bibinfo{pages}{527–535}.
\newblock


\bibitem[\protect\citeauthoryear{Weidlich and Wilkie}{Weidlich and
  Wilkie}{2007}]%
        {Weidlich:2007:layering}
\bibfield{author}{\bibinfo{person}{Andrea Weidlich} {and}
  \bibinfo{person}{Alexander Wilkie}.} \bibinfo{year}{2007}\natexlab{}.
\newblock \showarticletitle{Arbitrarily Layered Micro-Facet Surfaces}. In
  \bibinfo{booktitle}{\emph{Proceedings of the 5th International Conference on
  Computer Graphics and Interactive Techniques in Australia and Southeast
  Asia}} (Perth, Australia) \emph{(\bibinfo{series}{GRAPHITE '07})}.
  \bibinfo{pages}{171–178}.
\newblock


\bibitem[\protect\citeauthoryear{Weier and Belcour}{Weier and Belcour}{2020}]%
        {WeierAndBelcour:2020:Anisotropic}
\bibfield{author}{\bibinfo{person}{Philippe Weier} {and}
  \bibinfo{person}{Laurent Belcour}.} \bibinfo{year}{2020}\natexlab{}.
\newblock \showarticletitle{Rendering Layered Materials with Anisotropic
  Interfaces}.
\newblock \bibinfo{journal}{\emph{Journal of Computer Graphics Techniques
  (JCGT)}} \bibinfo{volume}{9}, \bibinfo{number}{2} (\bibinfo{date}{20 June}
  \bibinfo{year}{2020}), \bibinfo{pages}{37--57}.
\newblock
\showISSN{2331-7418}
\urldef\tempurl%
\url{http://jcgt.org/published/0009/02/03/}
\showURL{%
\tempurl}


\bibitem[\protect\citeauthoryear{Weinmann, Gall, and Klein}{Weinmann
  et~al\mbox{.}}{2014a}]%
        {Weinmann:2014:PCA}
\bibfield{author}{\bibinfo{person}{Michael Weinmann}, \bibinfo{person}{Juergen
  Gall}, {and} \bibinfo{person}{Reinhard Klein}.}
  \bibinfo{year}{2014}\natexlab{a}.
\newblock \showarticletitle{Material Classification Based on Training Data
  Synthesized Using a BTF Database}. In \bibinfo{booktitle}{\emph{Computer
  Vision -- ECCV 2014}}, \bibfield{editor}{\bibinfo{person}{David Fleet},
  \bibinfo{person}{Tomas Pajdla}, \bibinfo{person}{Bernt Schiele}, {and}
  \bibinfo{person}{Tinne Tuytelaars}} (Eds.). \bibinfo{publisher}{Springer
  International Publishing}, \bibinfo{address}{Cham},
  \bibinfo{pages}{156--171}.
\newblock


\bibitem[\protect\citeauthoryear{Weinmann, Gall, and Klein}{Weinmann
  et~al\mbox{.}}{2014b}]%
        {weinmann-2014-materialclassification}
\bibfield{author}{\bibinfo{person}{Michael Weinmann}, \bibinfo{person}{Juergen
  Gall}, {and} \bibinfo{person}{Reinhard Klein}.}
  \bibinfo{year}{2014}\natexlab{b}.
\newblock \showarticletitle{Material Classification Based on Training Data
  Synthesized Using a BTF Database}. In \bibinfo{booktitle}{\emph{Computer
  Vision - ECCV 2014 - 13th European Conference, Zurich, Switzerland, September
  6-12, 2014, Proceedings, Part III}}. \bibinfo{publisher}{Springer
  International Publishing}, \bibinfo{pages}{156--171}.
\newblock


\bibitem[\protect\citeauthoryear{Xia, Walter, Hery, and Marschner}{Xia
  et~al\mbox{.}}{2020}]%
        {Xia:2020:Layered}
\bibfield{author}{\bibinfo{person}{Mengqi~(Mandy) Xia}, \bibinfo{person}{Bruce
  Walter}, \bibinfo{person}{Christophe Hery}, {and} \bibinfo{person}{Steve
  Marschner}.} \bibinfo{year}{2020}\natexlab{}.
\newblock \showarticletitle{Gaussian Product Sampling for Rendering Layered
  Materials}.
\newblock \bibinfo{journal}{\emph{Computer Graphics Forum}}
  \bibinfo{volume}{39}, \bibinfo{number}{1} (\bibinfo{year}{2020}),
  \bibinfo{pages}{420--435}.
\newblock


\bibitem[\protect\citeauthoryear{Yamaguchi, Yatagawa, Tokuyoshi, and
  Morishima}{Yamaguchi et~al\mbox{.}}{2019}]%
        {Yamaguchi:2019:anisotropic}
\bibfield{author}{\bibinfo{person}{Tomoya Yamaguchi}, \bibinfo{person}{Tatsuya
  Yatagawa}, \bibinfo{person}{Yusuke Tokuyoshi}, {and} \bibinfo{person}{Shigeo
  Morishima}.} \bibinfo{year}{2019}\natexlab{}.
\newblock \showarticletitle{Real-Time Rendering of Layered Materials with
  Anisotropic Normal Distributions}. In \bibinfo{booktitle}{\emph{SIGGRAPH Asia
  2019 Technical Briefs}} \emph{(\bibinfo{series}{SA '19})}.
  \bibinfo{publisher}{Association for Computing Machinery},
  \bibinfo{address}{New York, NY, USA}, \bibinfo{pages}{87–90}.
\newblock
\urldef\tempurl%
\url{https://doi.org/10.1145/3355088.3365165}
\showDOI{\tempurl}


\bibitem[\protect\citeauthoryear{Yan, Sun, Jensen, and Ramamoorthi}{Yan
  et~al\mbox{.}}{2017}]%
        {yan2017furbssrdf}
\bibfield{author}{\bibinfo{person}{Ling-Qi Yan}, \bibinfo{person}{Weilun Sun},
  \bibinfo{person}{Henrik~Wann Jensen}, {and} \bibinfo{person}{Ravi
  Ramamoorthi}.} \bibinfo{year}{2017}\natexlab{}.
\newblock \showarticletitle{A BSSRDF Model for Efficient Rendering of Fur with
  Global Illumination}.
\newblock \bibinfo{journal}{\emph{ACM Transactions on Graphics (Proceedings of
  SIGGRAPH Asia 2017)}} \bibinfo{volume}{36}, \bibinfo{number}{6}
  (\bibinfo{year}{2017}).
\newblock


\bibitem[\protect\citeauthoryear{Zeltner and Jakob}{Zeltner and Jakob}{2018}]%
        {Zeltner2018Layer}
\bibfield{author}{\bibinfo{person}{Tizian Zeltner} {and}
  \bibinfo{person}{Wenzel Jakob}.} \bibinfo{year}{2018}\natexlab{}.
\newblock \showarticletitle{The Layer Laboratory: A Calculus for Additive and
  Subtractive Composition of Anisotropic Surface Reflectance}.
\newblock \bibinfo{journal}{\emph{Transactions on Graphics (Proceedings of
  SIGGRAPH)}} \bibinfo{volume}{37}, \bibinfo{number}{4} (\bibinfo{date}{July}
  \bibinfo{year}{2018}), \bibinfo{pages}{74:1--74:14}.
\newblock


\bibitem[\protect\citeauthoryear{Zhu, Bai, Xu, Bako,
  Vel\'{a}zquez-Armend\'{a}riz, Wang, Sen, Ha\v{s}an, and Yan}{Zhu
  et~al\mbox{.}}{2021}]%
        {Zhu:2021:Luminaires}
\bibfield{author}{\bibinfo{person}{Junqiu Zhu}, \bibinfo{person}{Yaoyi Bai},
  \bibinfo{person}{Zilin Xu}, \bibinfo{person}{Steve Bako},
  \bibinfo{person}{Edgar Vel\'{a}zquez-Armend\'{a}riz}, \bibinfo{person}{Lu
  Wang}, \bibinfo{person}{Pradeep Sen}, \bibinfo{person}{Milo\v{s} Ha\v{s}an},
  {and} \bibinfo{person}{Ling-Qi Yan}.} \bibinfo{year}{2021}\natexlab{}.
\newblock \showarticletitle{Neural Complex Luminaires: Representation and
  Rendering}.
\newblock \bibinfo{journal}{\emph{ACM Trans. Graph.}} \bibinfo{volume}{40},
  \bibinfo{number}{4}, Article \bibinfo{articleno}{57} (\bibinfo{date}{July}
  \bibinfo{year}{2021}), \bibinfo{numpages}{12}~pages.
\newblock


\end{thebibliography}

\end{document}